\documentclass[fleqn,usenatbib,useAMS]{mnras}

\usepackage{newtxtext,newtxmath}

\usepackage[T1]{fontenc}
\usepackage{ae,aecompl}
\usepackage{multirow}
\usepackage{rotating}
\usepackage{lscape} 
\usepackage{longtable}

\usepackage{graphicx}	
\usepackage{amsmath}	
\usepackage{amssymb}	
\usepackage{xcolor}

\title[ORBYTS: Ephemeris Refinement of Transiting Exoplanets]{Original Research By Young Twinkle Students (ORBYTS): Ephemeris Refinement of Transiting Exoplanets}

\author[B. Edwards et al.]{Billy Edwards$^{1}$\thanks{E-mail: \href{billy.edwards.16@ucl.ac.uk}},
Quentin Changeat$^{1}$,
Kai Hou Yip$^{1}$,
Angelos Tsiaras$^{1}$,
Jake Taylor$^{2}$, \newauthor
Bilal Akhtar$^{3}$,
Josef AlDaghir$^{3}$,
Pranup Bhattarai$^{3}$,
Tushar Bhudia$^{4}$,
Aashish \newauthor Chapagai$^{3}$,
Michael Huang$^{3}$,
Danyaal Kabir$^{4}$,
Vieran Khag$^{4}$,
Summyyah Khaliq$^{4}$, \newauthor
Kush Khatri$^{3}$,
Jaidev Kneth$^{4}$,
Manisha Kothari$^{4}$,
Ibrahim Najmudin$^{3}$,
Lobanaa \newauthor Panchalingam$^{4}$,
Manthan Patel$^{3}$,
Luxshan Premachandran$^{4}$,
Adam Qayyum$^{4}$,
Prasen \newauthor Rana$^{3}$,
Zain Shaikh$^{3}$,
Sheryar Syed$^{4}$,
Harnam Theti$^{4}$,
Mahmoud Zaidani$^{3}$,
Manasvee \newauthor Saraf$^{1}$,
Damien de Mijolla$^{1}$,
Hamish Caines$^{1}$,
Anatasia Kokori$^{5,6}$,
Marco Rocchetto$^{7,1}$, \newauthor
Matthias Mallonn$^{8}$,
Matthieu Bachschmidt$^{9}$,
Josep M. Bosch$^{10}$,
Marc Bretton$^{11}$, \newauthor
Philippe Chatelain$^{12}$,
Marc Deldem$^{13}$,
Romina Di Sisto$^{14,15}$,
Phil Evans$^{16,17}$,
Eduardo \newauthor Fern\'andez-Laj\'us$^{14,15}$,
Pere Guerra$^{18}$,
Ferran Grau Horta$^{19}$,
Wonseok Kang$^{20}$,
Taewoo \newauthor Kim$^{20}$,
Arnaud Leroy$^{21}$,
Franti\v{s}ek Lomoz$^{22}$,
Juan Lozano de Haro$^{23}$,
Veli-Pekka \newauthor Hentunen$^{24}$,
Yves Jongen$^{25}$,
David Molina$^{26}$,
Romain Montaigut$^{21}$,
Ramon Naves$^{27}$,  \newauthor
Manfred Raetz$^{28}$,
Thomas Sauer$^{29}$,
Americo Watkins$^{30}$,
Ana\"el W\"unsche$^{11}$,
Martin \newauthor Zibar$^{31}$,
William Dunn$^1$,
Marcell Tessenyi$^{32,1}$,
Giorgio Savini$^{1,33,32}$, 
Giovanna \newauthor Tinetti$^{1,32}$ \& 
Jonathan Tennyson$^{1,32}$
\\
}
\date{}

\pubyear{2019}

\begin{document}
\label{firstpage}
\pagerange{\pageref{firstpage}--\pageref{lastpage}}
\maketitle

\begin{abstract}

We report follow-up observations of transiting exoplanets that have either large uncertainties (\textgreater 10 minutes) in their transit times or have not been observed for over three years. A fully robotic ground-based telescope network, observations from citizen astronomers and data from TESS have been used to study eight planets, refining their ephemeris and orbital data. Such follow-up observations are key for ensuring accurate transit times for upcoming ground and space-based telescopes which may seek to characterise the atmospheres of these planets. We find deviations from the expected transit time for all planets, with transits occurring outside the 1 sigma uncertainties for seven planets. Using the newly acquired observations, we subsequently refine their periods and reduce the current predicted ephemeris uncertainties to 0.28 - 4.01 minutes. A significant portion of this work has been completed by students at two high schools in London as part of the Original Research By Young Twinkle Students (ORBYTS) programme.


\end{abstract}

\section{Introduction}

Over the past decade, thousands of exoplanets have been detected via the transit method. Current and future observatories such as the Transiting Exoplanet Survey Satellite (TESS, \cite{ricker}) and the PLAnetary Transits and Oscillations of stars (PLATO, \cite{rauer}) satellite are expected to discover tens of thousands more. Upcoming ground and space-based telescopes such as the European Extremely Large Telescope (E-ELT, \cite{brandl}), the Thirty Meter Telescope (TMT, \cite{skidmore}), the Giant Magellan Telescope (GMT, \cite{fanson}), the James Webb Space Telescope (JWST, \cite{greene}), Twinkle \citep{edwards_twinkle} and Ariel \citep{tinetti_ariel} will characterise the atmospheres of a large population of exoplanets via transit and eclipse spectroscopy at visible and infrared wavelengths. These missions will move the exoplanet field from an era of detection into one of characterisation, allowing for the identification of the molecular species present and their chemical profile, insights into the atmospheric temperature profile and the detection and characterisation of clouds (e.g. \cite{rocchetto,rodler,yui_clouds,changeat}). However, observing time on these exceptional facilities will be precious. Therefore observations of transiting exoplanets will need to have a limited time window while ensuring that enough margin is included to avoid a transit event being partially, or completely, missed. Large errors in the ephemeris of a planet increase the observation time required to ensure the full transit is captured and thus reduce the efficiency, and science yield, of these missions. Accurate ephemeris data, collected over a long baseline, can also be used to search for, and characterise, other planets in the system via transit time variations.

Just after discovery, the time of the next transit for a planet is well known. Unfortunately the accuracy of predicted future transits degrades over time due to the increased number of epochs since the last observation and the stacking of the period error. In extreme cases this can mean the transit time is practically lost, with errors of several hours (e.g. Corot-24 b \& c, \cite{alonso}). In addition to this, extrapolating transit times from only a few data points over a limited baseline can easily introduce bias (e.g. \cite{benneke}). Finally, we could expect transit times to shift due to dynamical phenomena such as tidal orbital decay, apsidal precession or from gravitational interactions with other bodies in the system (see e.g. \cite{agol, maciejewski, bouma}). These can only be understood, and mitigated for, by regularly observing targets over a long baseline. In the era of TESS, which is expected to find several thousand transiting planets \citep{sullivan,barclay}, this will become increasingly difficult due to the sheer number of targets and require a coordinated effort by many groups and telescope networks to prepare for characterisation by the next generation facilities. This campaign will need data from both ground-based facilities and space-based telescopes such as TESS, CHEOPS and Twinkle.

Ground-based follow-up will require not only a large number of telescopes but many person-hours to plan observations and process the data. Citizen astronomers, citizen science and educational outreach offer an excellent opportunity to support future space missions. Given the brightness of the host stars of planets found by TESS, even modestly sized telescopes can be used to re-observe these systems, reducing the errors on their ephemeris. The ability of small ground-based telescopes to contribute to exoplanet science is well known (e.g. \cite{kabath}). The Next Generation Transit Survey (NGTS, \cite{wheatley}) has shown that sub-millimag precision is achievable by simultaneously observing the same transit with many identical small telescopes and combining data. Using such methods could expand the number of exoplanets that are observable from the ground.

Here we present a project to refine the ephemerides of eight exoplanets using a fully robotic telescope network, observations from citizen astronomers and data from TESS; a significant portion of the work has been completed by high school students via the ORBYTS programme.

\section{Outreach and Citizen Science Projects}

The observations and analysis presented here have been achieved via engagement with several citizen science and outreach initiatives and a brief description of each of these is given below.

\subsection{ORBYTS}

Original Research By Young Twinkle Students (ORBYTS) is an educational programme in which secondary school pupils work on original research linked to the Twinkle Space Mission under the tuition of PhD students and other young scientists \citep{mckemmish_orbyts,sousa-silva}. The ORBYTS programme has been run since 2012 and is jointly managed by Blue Skies Space Ltd. (BSSL) and University College London (UCL). The Twinkle Space Mission\footnote{\url{http://www.twinkle-spacemission.co.uk}} is a new, fast-track satellite designed for launch by 2024 and has been conceived for providing faster access to space-based spectroscopic data. While the satellite will also survey Solar System objects \citep{edwards_mss, edwards_sbs}, a key science case for Twinkle is the characterisation of a population of extrasolar planets via transit and eclipse spectroscopy \citep{edwards_twinkle}. ORBYTS offers school pupils the chance to enrich our understanding of these new worlds by improving our knowledge of the molecules they're made of, their orbits and their physical properties. This provides a unique opportunity for pupils to undertake cutting-edge science that has a meaningful impact on a future space mission.

To achieve this, ORBYTS partners dynamic, passionate science researchers with secondary schools, where, through fortnightly school visits over an academic year, the researcher teaches the students undergraduate-level physics. The goal of every partnership is that school students will have the opportunity to use this new knowledge to contribute towards publishable research. Pupils get hands on experience of scientific research and work closely with young scientists. By partnering schools with relatable researchers, the programme aims to not only improve student aspirations and scientific literacy, but also help to address diversity challenges by dispelling harmful stereotypes, challenging any preconceptions about who can become a scientist. The organisers and tutors strongly believe that all school students should have the opportunity to become involved in active scientific research and to be culturally connected to space missions. Previous projects have included providing accurate molecular transition frequencies with the ExoMol group \citep{chubb,chubb_2,jt672,mckemmish_marvel,jt764} as these line lists are crucial for atmospheric retrievals. 

During the current project the students selected suitable follow-up targets, scheduled observations and analysed the observational data.

\subsection{Exoplanet Transit Database}

The Exoplanet Transit Database (ETD, \cite{poddany}) was established in 2008 and is a web-based application which is open to any exoplanet observer. The ETD is a project of the Variable Star and Exoplanet Section of the Czech Astronomical Society and the site consists of three parts, the first of which provides predictions of the upcoming transits. The second section allows for users to upload new data and the final function is the display of the observed - calculated diagrams (O-C). The ETD has hundreds of contributors and the database contains thousands of observations. While all observations are analysed by the ETD system to produce these graphs, the data can also be downloaded. The ETD does not facilitate a ranking of planets based on their current uncertainties.

\subsection{ExoWorlds Spies}

ExoWorlds Spies\footnote{\url{https://exoworldsspies.com}} is a project that started in early 2018, aiming to monitor transiting exoplanets through long-term regular observations using small and medium scale telescopes. This effort, is supported by citizen astronomers, the Holomon Astronomical Station and the Telescope Live network. The project promotes the idea that research is an effort that everyone can contribute and, thus, it is open to collaborations with the public, including school and university students. To facilitate this, user-friendly data analysis tools and a dedicated website have been developed as part of the project, in order to disseminate the material to as many people as possible. The website includes audiovisual material, information on the project, data analysis tools, instructions, observational data and graphics. All sources are online, free, and available for everyone both in English and Greek. So far, the ExoWorlds Spies database includes approximately 60 transit observations of more than 25 different exoplanets, from both the North and the South hemisphere, including recently discovered planets with limited data available. A number of these transits are already available on the website for members of the general public, students, and citizens to analyse.

\subsection{ExoClock}

ExoClock has been established as part of the ground-based characterisation campaign for the Ariel space mission. Ariel aims to observe 1000 exoplanets during its primary mission, characterising their atmospheres and seeking to understand the chemical diversity of planets in our galaxy \citep{tinetti_ariel}. The Ariel Mission Reference Sample (MRS), the planets observed by the mission, will be selected from a large list of potential targets. The selection criterion will aim to produce a multifarious population of planets for study. However, the lack of basic knowledge such as stellar variability and the expected transit time of the system, may mean a planet is not selected for observation, potentially reducing the impact of the mission. ExoClock aims to facilitate a coordinated programme of ground-based observations to maximise the efficiency of the Ariel mission. The programme also aims to stimulate engagement with citizen astronomers, allowing them to contribute to an upcoming ESA mission. The site ranks the potential Ariel targets from \cite{edwards_ariel}, prioritising those that have a large uncertainty in their next transit time. These can then be filtered by the location of the observer and the telescope size, providing a list of exoplanet transits which would be observable in the near future. The ExoClock initiative has the explicit rule that all those who upload data for a planetary system will be included on any subsequent publications. 

\section{Target Selection}

\begin{table*}
    \centering
    \caption[Data on exoplanets for which observations were acquired]{Exoplanets for which observations were acquired and the calculated uncertainty in their transit mid time on 1$^{st}$ July 2019 based on data from the NASA Exoplanet Archive.}
    \resizebox{0.95\textwidth}{!}{%
    \begin{tabular}{lccrcl}\hline\hline
     Planet & Planet Radius [R$_J$] & Star V Mag & Uncertainty [minutes] & Last Observed & Reference \\\hline
     CoRoT-6 b & 1.17 & 13.9 & 2.7 &  2010 &\cite{fridlund}\\
     KELT-15 b & 1.44 & 11.2 & 15.7 & 2015 & \cite{rodriguez}\\
     KPS-1 b & 1.03 & 13.0 & 57.6 & 2018 & \cite{burdanov}\\
     K2-237 b & 1.65 & 11.6 & 13.0$^{\star}$ & 2018 & \cite{soto}\\
     WASP-45 b & 1.16 & 12.0 & 5.3 & 2013 & \cite{anderson}\\
     WASP-83 b & 1.04 & 12.9 & 11.9 & 2015 & \cite{hellier}\\
     WASP-119 b & 1.40 & 12.2 & 15.7 & 2016 & \cite{maxted}\\
     WASP-122 b & 1.74 & 11.0 & 4.9$^{\dagger}$ & 2016 & \cite{turner} \\\hline\hline
     \multicolumn{5}{l}{$^{\star}$The independent discovery paper \citep{smith} suggests an uncertainty 3.8 minutes}\\
    \multicolumn{5}{l}{$^{\dagger}$The independent discovery paper \citep{rodriguez} suggests an uncertainty 3.4 minutes}
    \end{tabular}
    }
 
    \label{tab:planets}
\end{table*}

There are several major exoplanet catalogues from which one can compile a list of potential planets. The most widely used and comprehensive is the NASA Exoplanet Archive\footnote{\url{https://exoplanetarchive.ipac.caltech.edu}} \citep{akeson}. The NASA catalogue was accessed in February 2019 and the transit error by mid 2019 (the end time of this project) was calculated for each planet. The next transit of a planet, $T_c$, can be calculated from
\begin{equation}
	T_c = T_0 + n \cdot P
\end{equation}
where $P$ is the period of the planet, $T_0$ is the last measured transit time and n is the number of epochs since this last observation. Both $T_0$ and $P$ have errors associated with their measurement and thus the error on the predicted transit time, $\Delta T_c$ is given by
\begin{equation}
	\Delta T_c = \sqrt{\Delta T_0^2 + (n \cdot \Delta P)^2}
\end{equation}
assuming no co-variance between the two parameters. There is, of course, a correlation between the fitted period and mid time but this co-variance is generally negligible. Suitable targets were found by filtering this list to include only those with a large transit uncertainty (\textgreater 10 minutes) or those that had not been observed for 3 or more years. We note that the ephemeris of many of the large, gaseous planets with significant transit uncertainties were refined by \cite{mallonn} and these were excluded from the study. The choice of targets was restricted by the size of the telescopes (0.35 -- 0.6 m, see Section \ref{data acq}) due to the star magnitude and transit depth but still many planets with substantial ephemeris errors were found to be observable. 

\section{Data Acquisition}
\label{data acq}

Table \ref{tab:planets} contains the planets for which data was obtained and the expected transit error on 1$^{st}$ July 2019. Although some of the planets observed here are around relatively fainter stars, they are all potentially suitable for spectroscopic follow-up and could be observed by Ariel \citep{edwards_ariel}. They may also be potential targets for characterisation by Twinkle, JWST or ground-based facilities. Observations of these targets were scheduled between February and April 2019.

\subsection{Robotic Ground-based Telescope Network}

For the new observations presented here we use the Telescope Live global network of robotic telescopes\footnote{\url{https://telescope.live}}. Telescope Live is a web application offering end-users the possibility to purchase images obtained on-demand from a network of robotic telescopes. Telescope Live kindly provided access to their telescopes for a total of 150 hours. At the time of writing the network consists of eight telescopes distributed across three observatories: a 1m ASA RC-1000AZ, a 0.6m Planewave CDK24 and two 0.5m ASA 500N located at El Sauce Observatory in Chile; a 0.7m ProRC 700 and two 0.1m Takahashi FSQ-106ED located at IC Astronomy in Spain; a 0.1m Takahashi FSQ-106ED located at Heaven's Mirror Observatory in Australia. The majority of observations were performed using a V filter (Luminance). Additionally we obtained a light curve of WASP-122 b using a 1.0 m Sinistro from the Las Cumbres Observatory (LCO) network\footnote{\url{https://lco.global}} thanks to the Educational Proposal FTPEPO2014A-004 led by Paul Roche.

\subsection{ETD and ExoWorldSpies}

For the selected planets, the ETD was searched for additional observations. The ETD provides a ranking of data quality from 1 to 5. Having removed observations with a data quality of less than 3, as well as excluding other unsuitable light curves via visual inspection, we found a total of 31 light curves from citizen astronomers: 5 of CoRoT-6 b, 21 of KPS-1 b, 3 of WASP-45 b and 2 of WASP-122 b. All these observations were undertaken as part of the TRansiting ExoplanetS and CAndidates (TRESCA) project\footnote{\url{http://var2.astro.cz/EN/tresca}} and are summarised in Table \ref{tab:all observations}. From ExoWorldSpies, we included an observation of WASP-83 b in our analysis. Additionally, the new observations taken as part of this work have been added to the ExoWorldsSpies and ExoClock databases.

\subsection{TESS}

Having observed the southern hemisphere, TESS is now surveying the northern hemisphere and thus has observed several of the planets studied here. We searched the Mikulski Archive for Space Telescopes (MAST\footnote{\url{http://archive.stsci.edu}}) and found that TESS has observed K2-237 b, KELT-15 b, WASP-45 b, WASP-83 b, WASP-119 b and WASP-122 b. A pipeline was built to find, acquire, reduce, and analyse the data. For a given target, the code searches MAST and returns all the data collected on the host star from various observatories. The list is filtered to see if TESS has observed the star and, if so, the Presearch Data Conditioning (PDC) light curve, which has had non-astrophysical variability removed and `bad data' eliminated through the methods outlined in the TESS guide\footnote{\url{https://spacetelescope.github.io/notebooks/notebooks/MAST/TESS/beginner_tour_lc_tp/beginner_tour_lc_tp.html}}, is downloaded. The data product is a time-series for each sector ($\sim$27 days) with a cadence of two minutes. After excluding the poor data, we recovered 7 K2-237 b transits, 12 for KELT-15 b, 8 for WASP-45 b, 4 of WASP-83 b along with 37 for WASP-119 b and 13 of WASP-122 b.

\section{Data Reduction and Analysis}

The Telescope Live network automatically gathers calibration frames and provides the data in a reduced format (though the raw and calibration frames can also be downloaded). These frames were analysed using the HOlomon Photometric Software (HOPS) which aligns the frames and normalises the flux of the target star by using selected comparison stars. This software is open-source and available on Github\footnote{\url{https://github.com/HolomonAstronomicalStation/hops}}.

The photometric light curves from all sources were fitted using PyLightcurve \citep{tsiaras_plc}, another code which is publicly available\footnote{\url{https://github.com/ucl-exoplanets/pylightcurve}}. Initially fit parameters were the orbital semi-major axis scaled by the stellar radius ($a/R_*$), the orbital inclination ($i$), the planet-star radius ratio ($R_p/R_*$), the midpoint of the transit ($T$) and the orbital period ($P$). In each case the Markov chain Monte Carlo (MCMC) was run with 100,000 iterations, a burn of 30,000 and 200 walkers. The limb darkening coefficients were fixed to theoretical values from \cite{claretI,claretII} according to the stellar parameters obtained from the planet discovery papers. Previous analyses show that the trends in ground-based light curves can be approximated with simple functions of only very few free parameters, for example low order polynomials over time (e.g. \cite{southworth,maciejewski,mackebrandt,mallonn}). Hence we detrended all ground-based light curves using a simple second-order polynomial. We then removed all data points with residuals greater than 3 sigma from the best-fit model. For TESS data, we used the flatten function from Wotan \citep{hippke}, an open-source python suite developed for comprehensive time-series detrending of exoplanet transit survey data\footnote{\url{https://github.com/hippke/wotan}}.


Next we fitted each light curve individually with $a/R_*$, $i$ and $R_p/R_*$ allowed to vary within 1 $\sigma$ of the values from the literature (or the new values computed here) while $T$ was fitted with bounds of 3$\sigma$. For targets which had been observed by TESS, we refined the planet transit parameters ($R_p/R_*$, $i$, $a/R_*$) and these are provided in Tables \ref{tab:para 1}-\ref{tab:para 3}. The uncertainties on each fitted mid-time are obtained from the posterior distributions of the MCMC chains. We convert all our mid-times into BJD$_{TDB}$ using the tool from \cite{eastman}. Having fit the mid transit time for all the data, we use a weighted least squares fit to obtain a linear period for the data analysed in this work and any previous mid-times from the literature (also converted to BJD$_{TDB}$). We varied the reference transit time, $T_0$, and report the value which minimised the co-variance between $T_0$ and $P$. 

\begin{figure}
    \centering
    \includegraphics[width = \columnwidth]{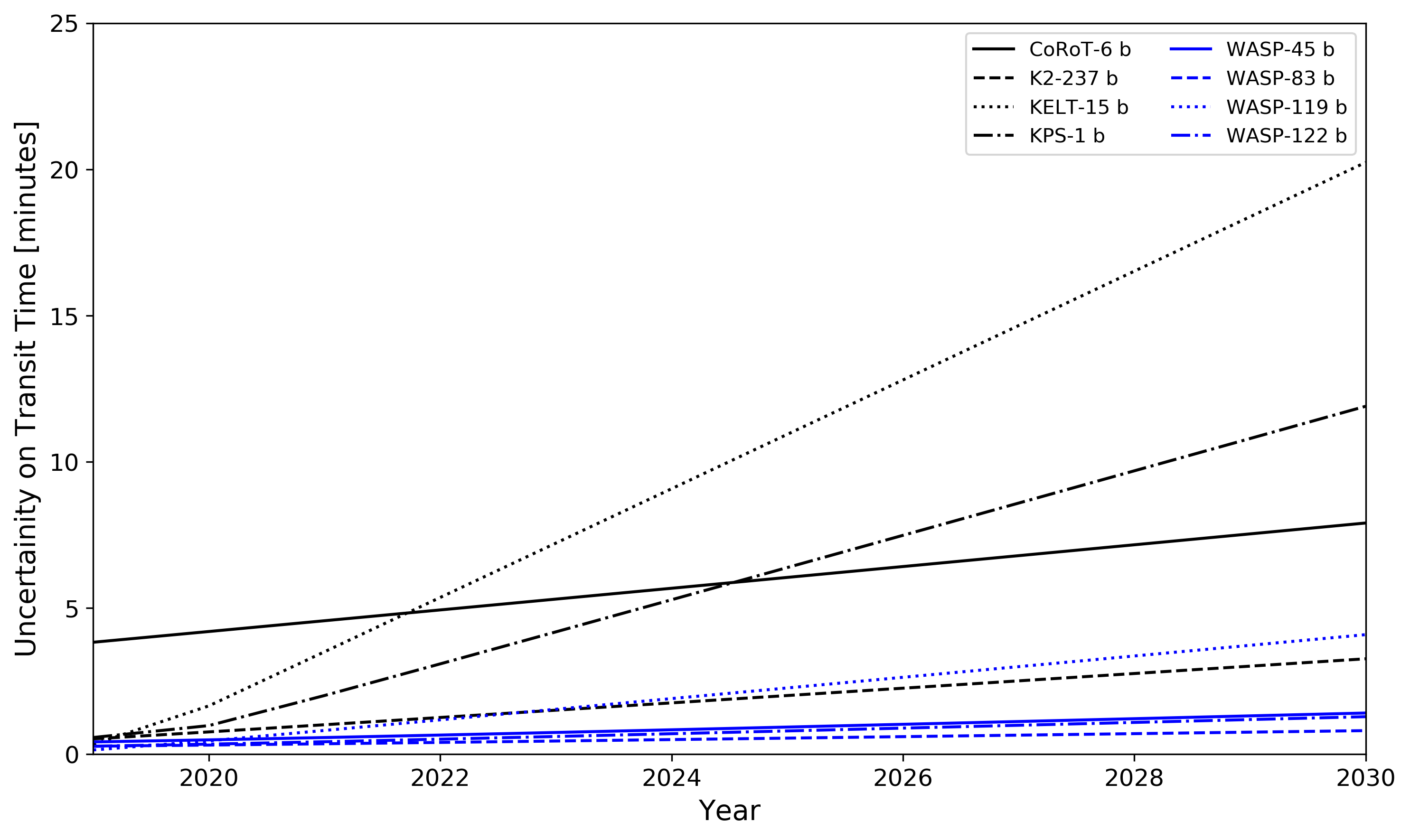}
    \caption{Projected uncertainties in the transit time of the planets studied here.}
    \label{future_uncert}
\end{figure}

Finally, we used the literature ephemerides from Table \ref{tab:lit_data} to compute `observed minus calculated' residuals for all transit times and used our refined ephemeris to predict the uncertainty on the transit times when Ariel is launched in 2028 as shown in Figure \ref{future_uncert}.

\begin{table*}
    \centering
    \caption{Summary of literature ephemeris data used here.}
    \begin{tabular}{cccc}\hline
      Planet & Mid Time [$BJD_{TDB}$] & Period [days] & Reference\\\hline
      CoRoT-6 b & 2454595.6144 $\pm$0.0002 & 8.886593$\pm$0.000004 & \cite{fridlund}\\
      K2-237 b & 2457684.8101 $\pm$ 0.0001 & 2.18056$\pm$0.00002 & \cite{soto}\\
      KELT-15 b & 2457029.1663$^{+0.0078}_{-0.0073}$ & 3.329441$\pm$0.000016 & \cite{rodriguez}\\
      KPS-1 b & 2457508.37019$^{+0.0079}_{-0.0078}$ & 1.706291$\pm$0.000059 & \cite{burdanov}\\
      WASP-45 b & 2455441.269995$\pm$0.00058 & 3.1260876$\pm$0.0000035 & \cite{anderson}\\
      WASP-83 b & 2455928.886085$\pm$0.0004 & 4.971252$\pm$0.000015 & \cite{hellier}\\
      WASP-119 b & 2456537.547779$\pm$0.002 & 2.49979$\pm$0.00001 & \cite{maxted}\\
      WASP-122 b & 2456665.224782$\pm$0.00021 & 1.7100566$^{+0.0000032}_{-0.0000026}$ & \cite{turner}\\\hline
    \end{tabular}
    \label{tab:lit_data}
\end{table*}

\section{Results}

\begin{figure*}
    \centering
    \includegraphics[width = 0.475\textwidth]{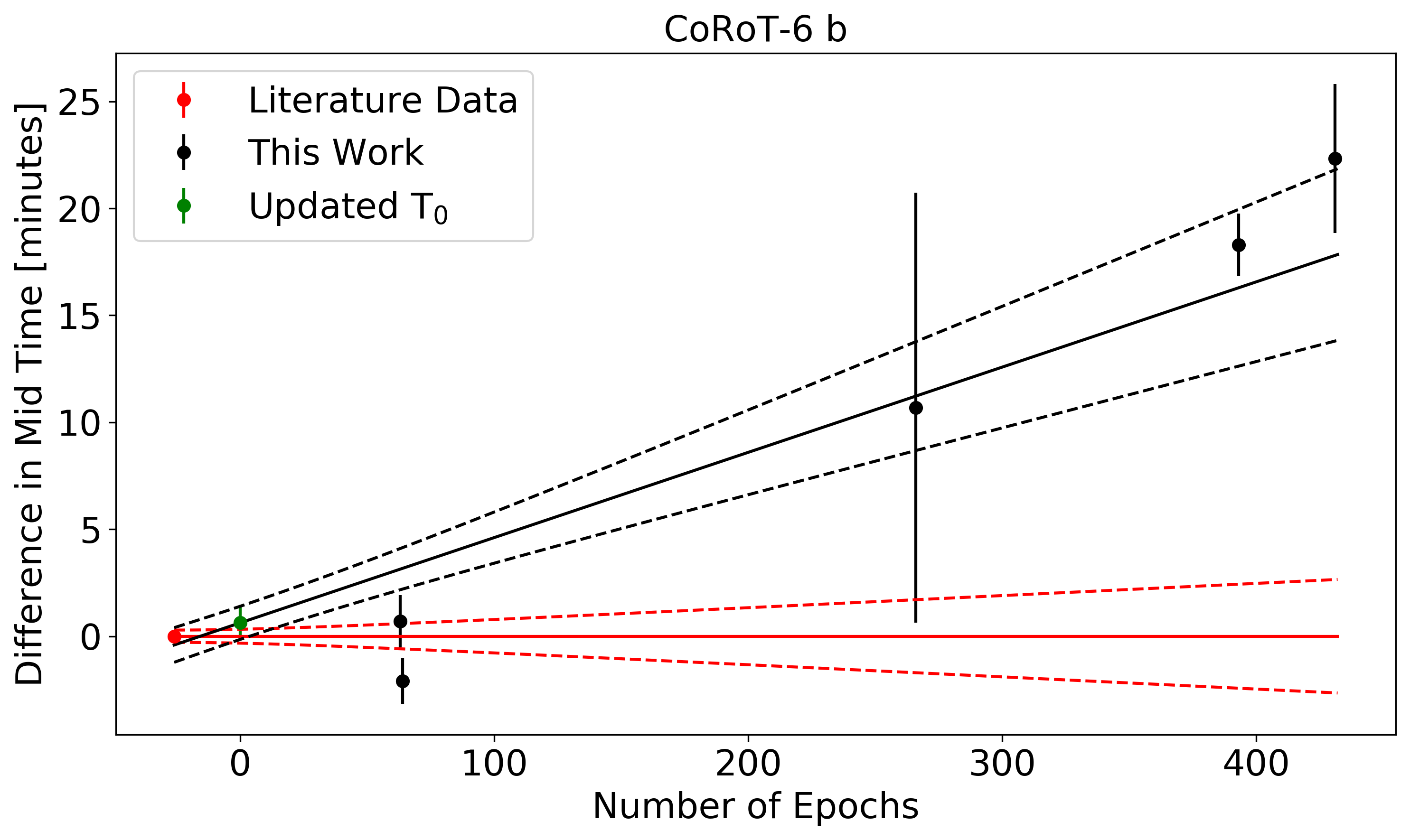}
    \includegraphics[width = 0.475\textwidth]{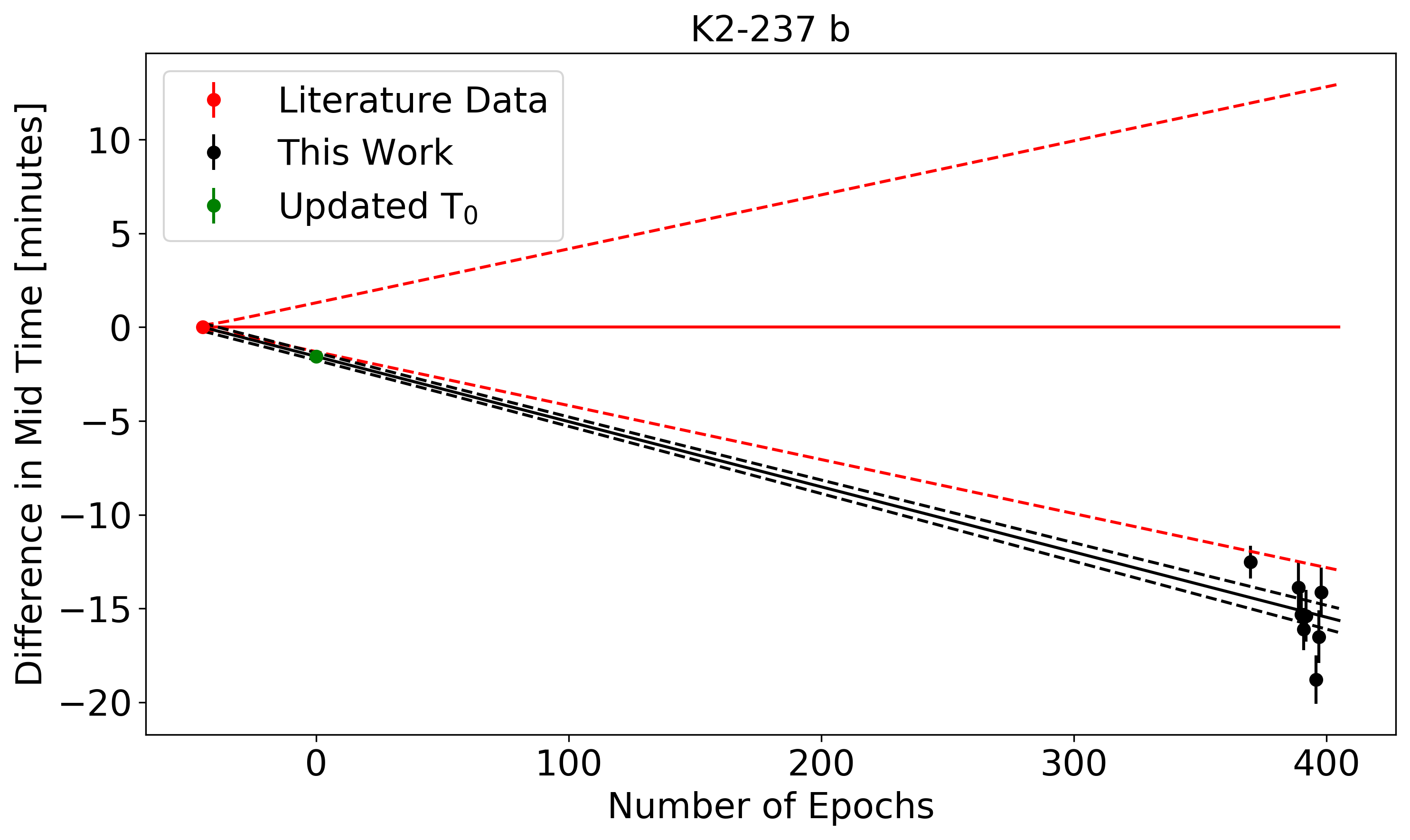}
    \includegraphics[width = 0.475\textwidth]{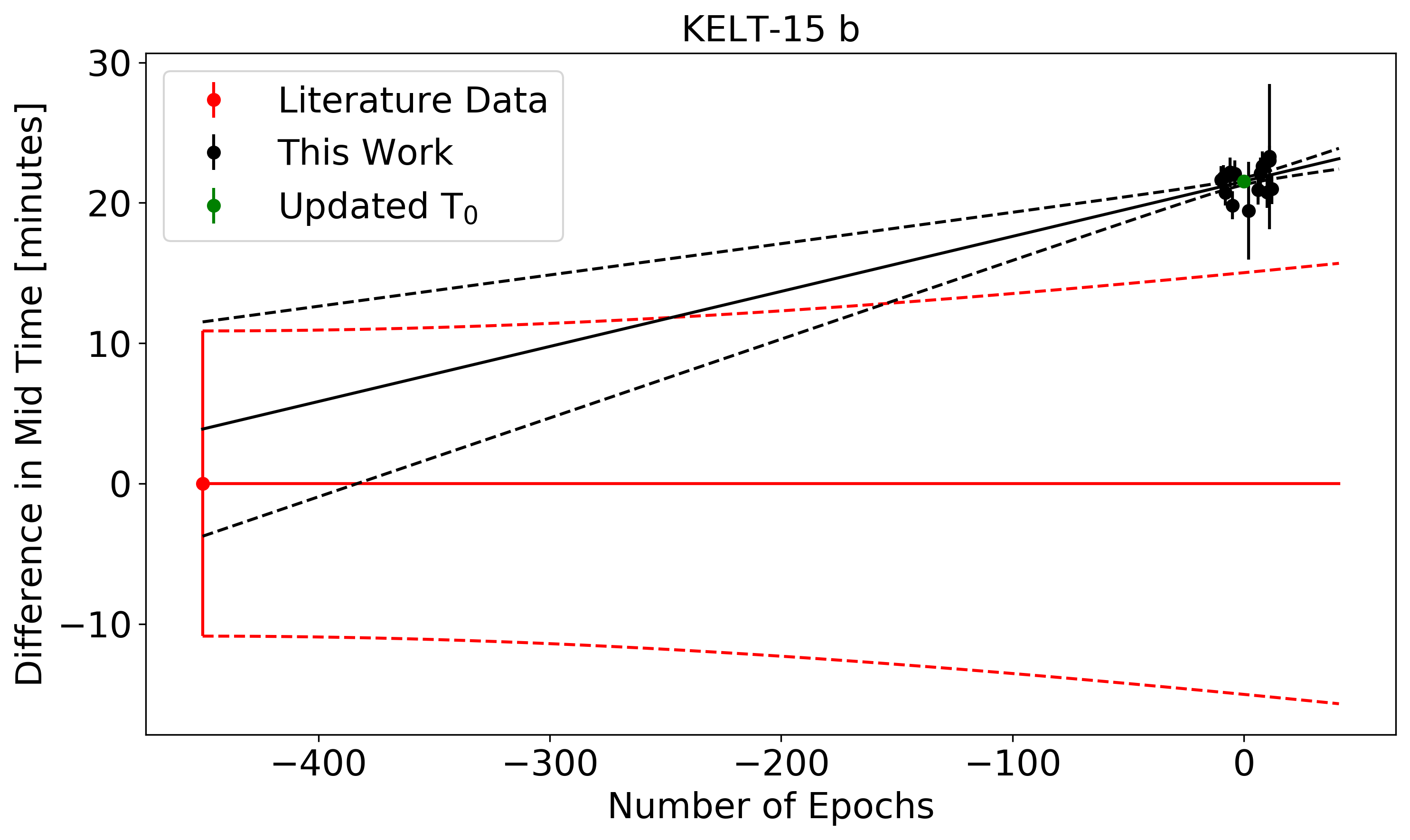}
    \includegraphics[width = 0.475\textwidth]{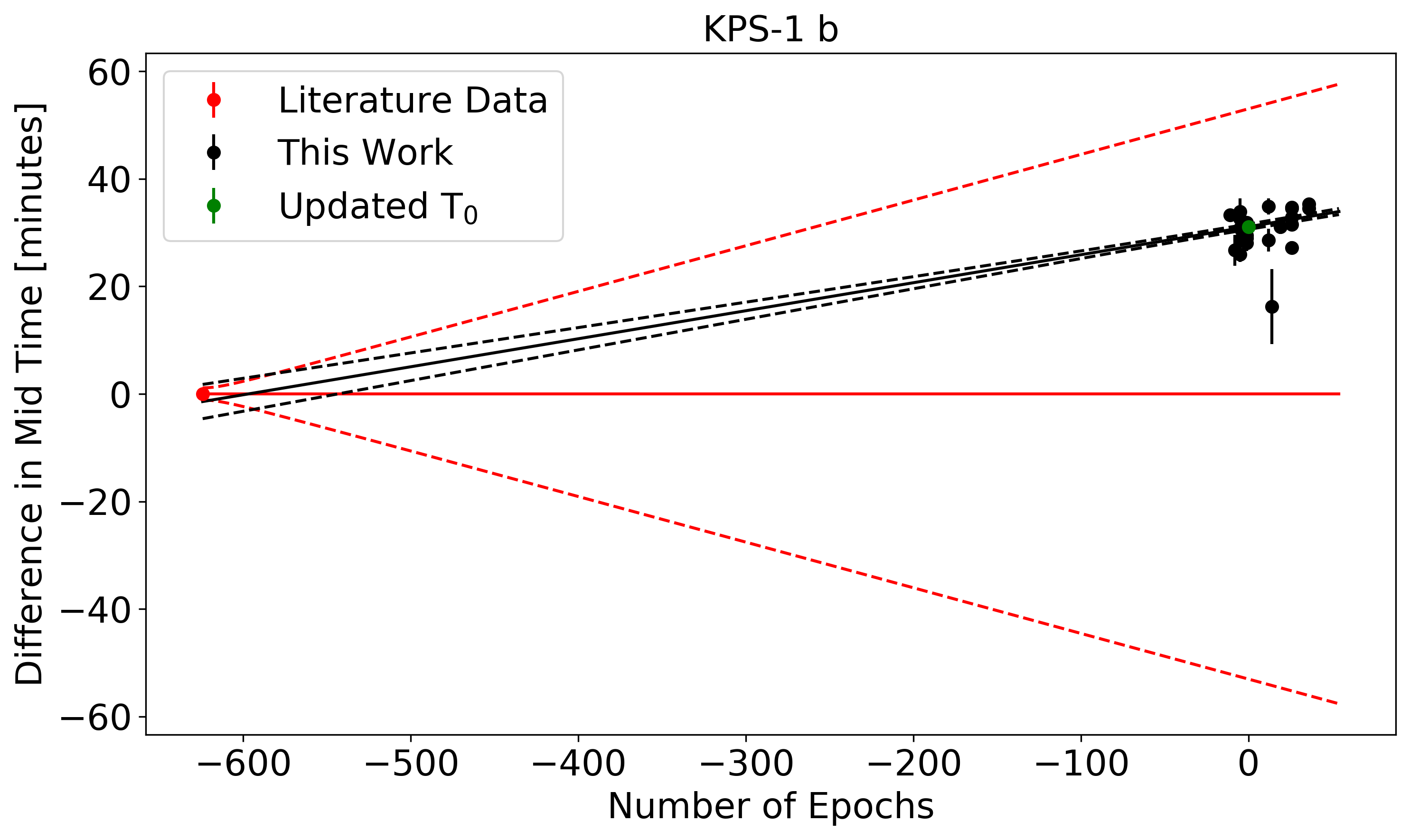}
    \includegraphics[width = 0.475\textwidth]{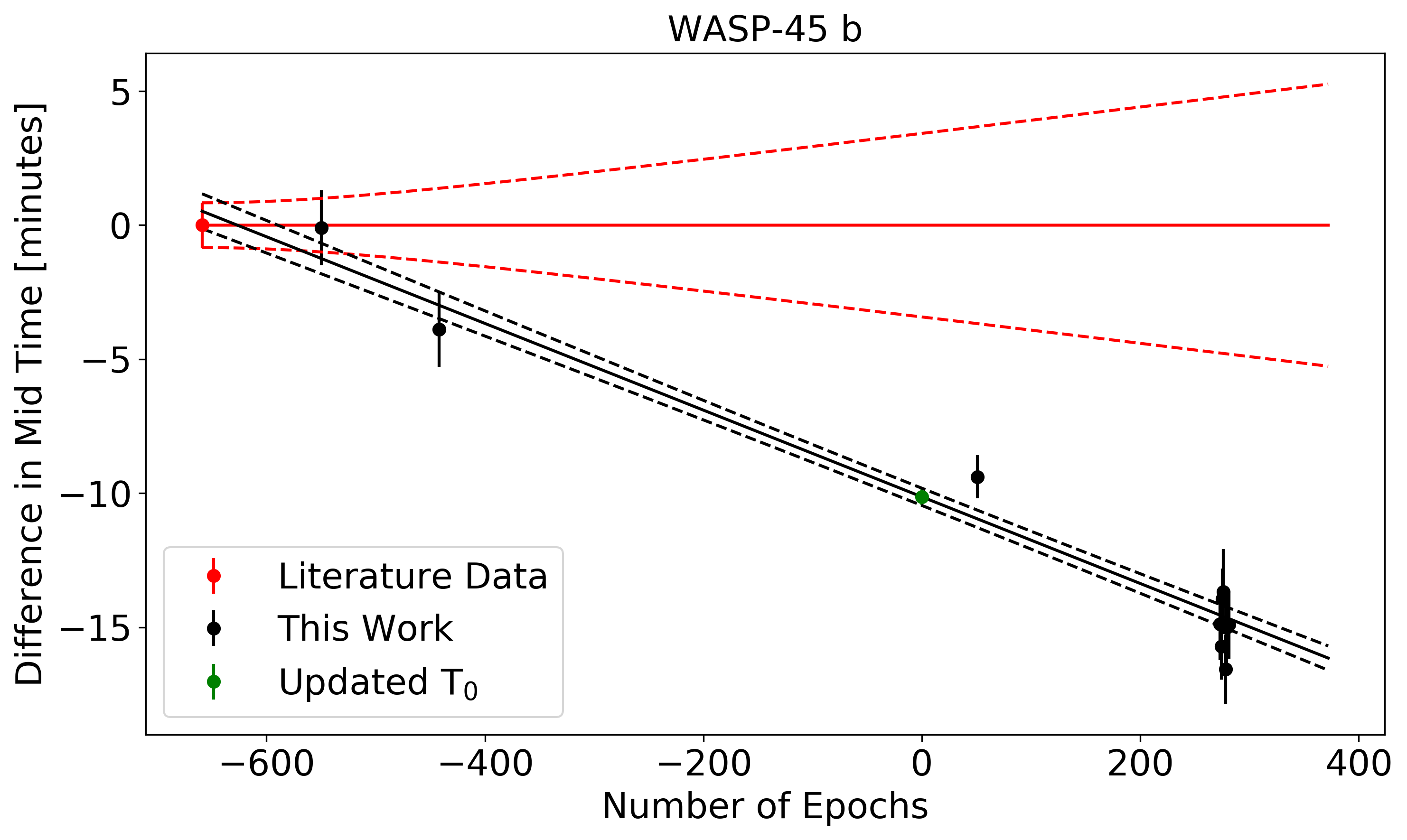}
    \includegraphics[width = 0.475\textwidth]{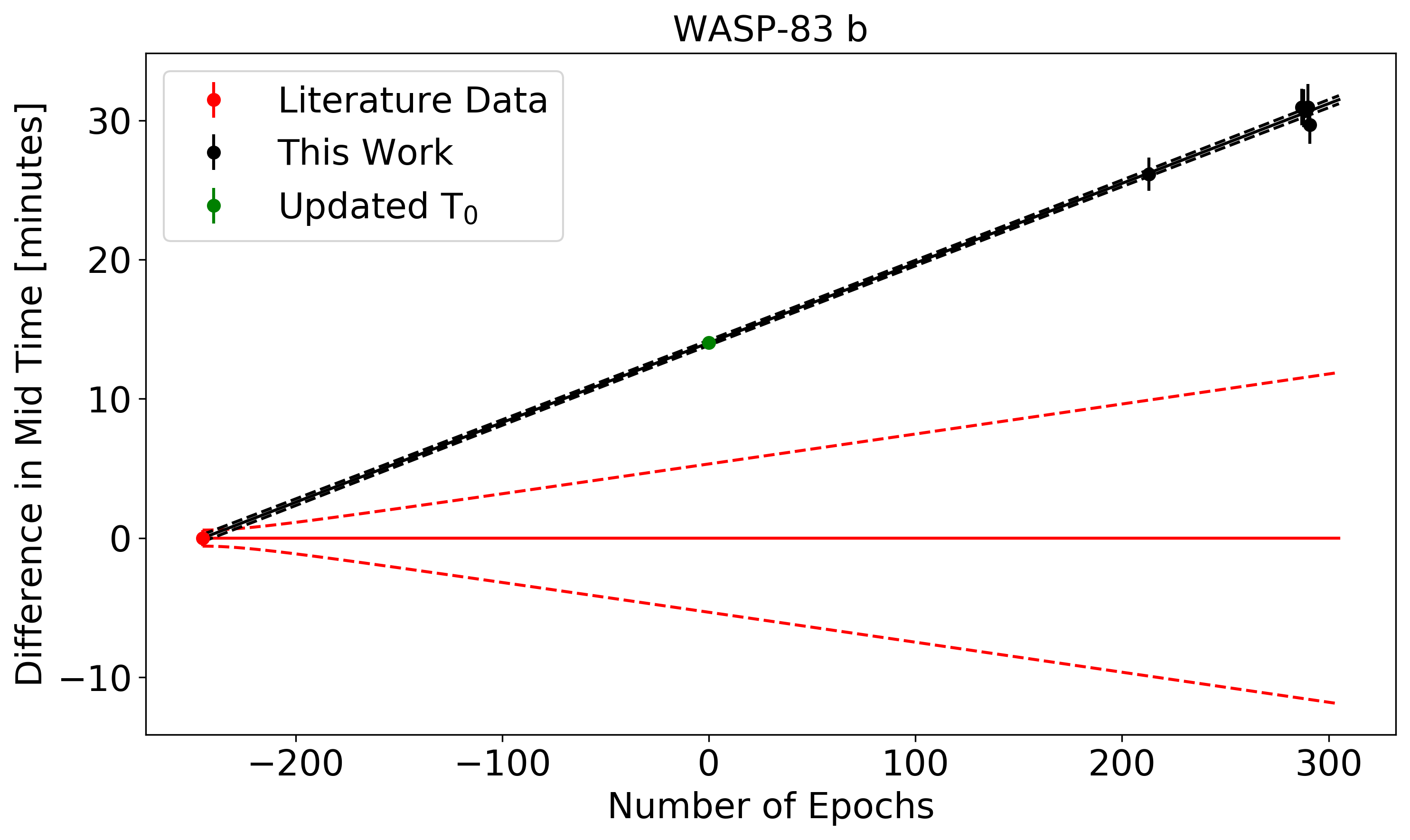}
    \includegraphics[width = 0.475\textwidth]{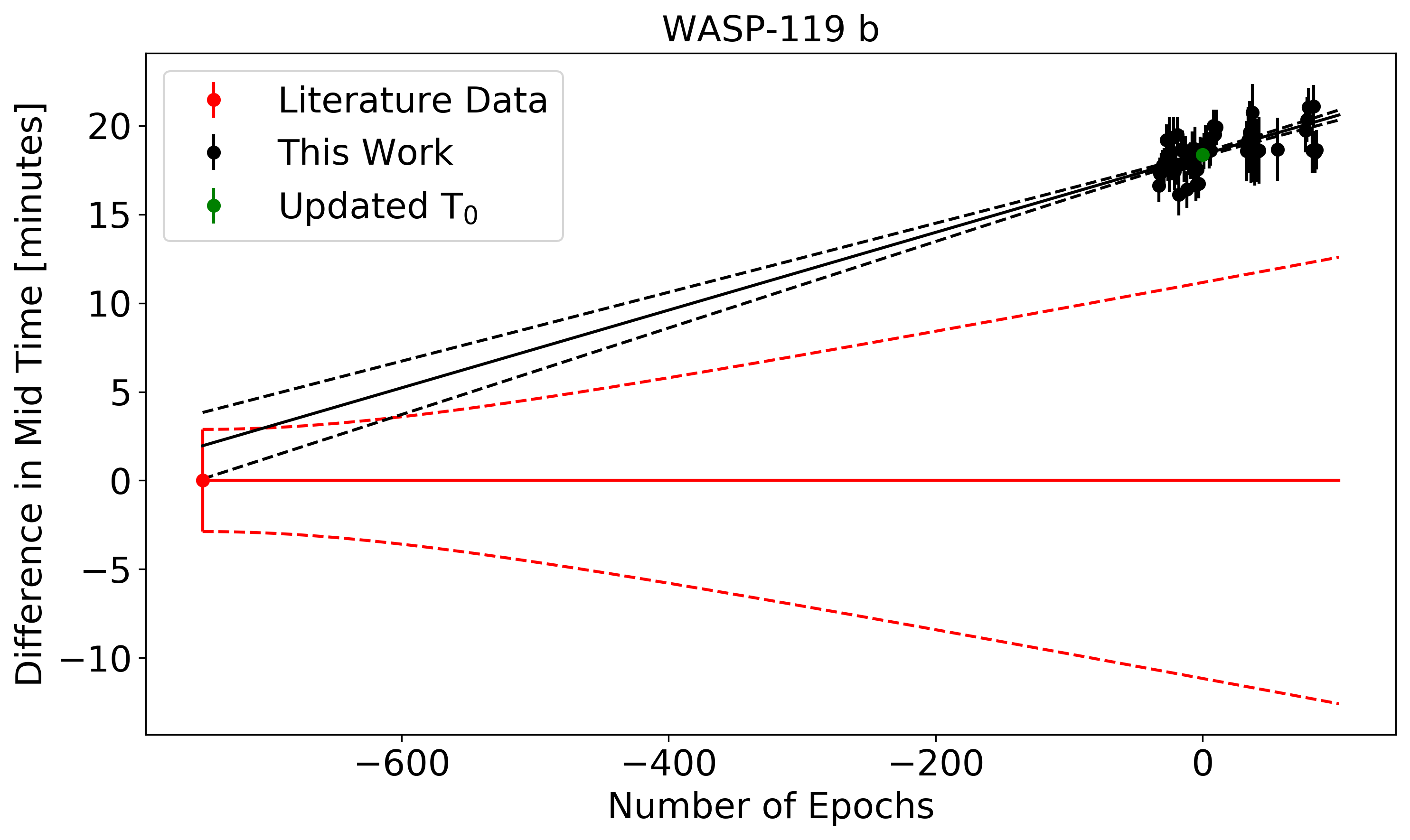}
    \includegraphics[width = 0.475\textwidth]{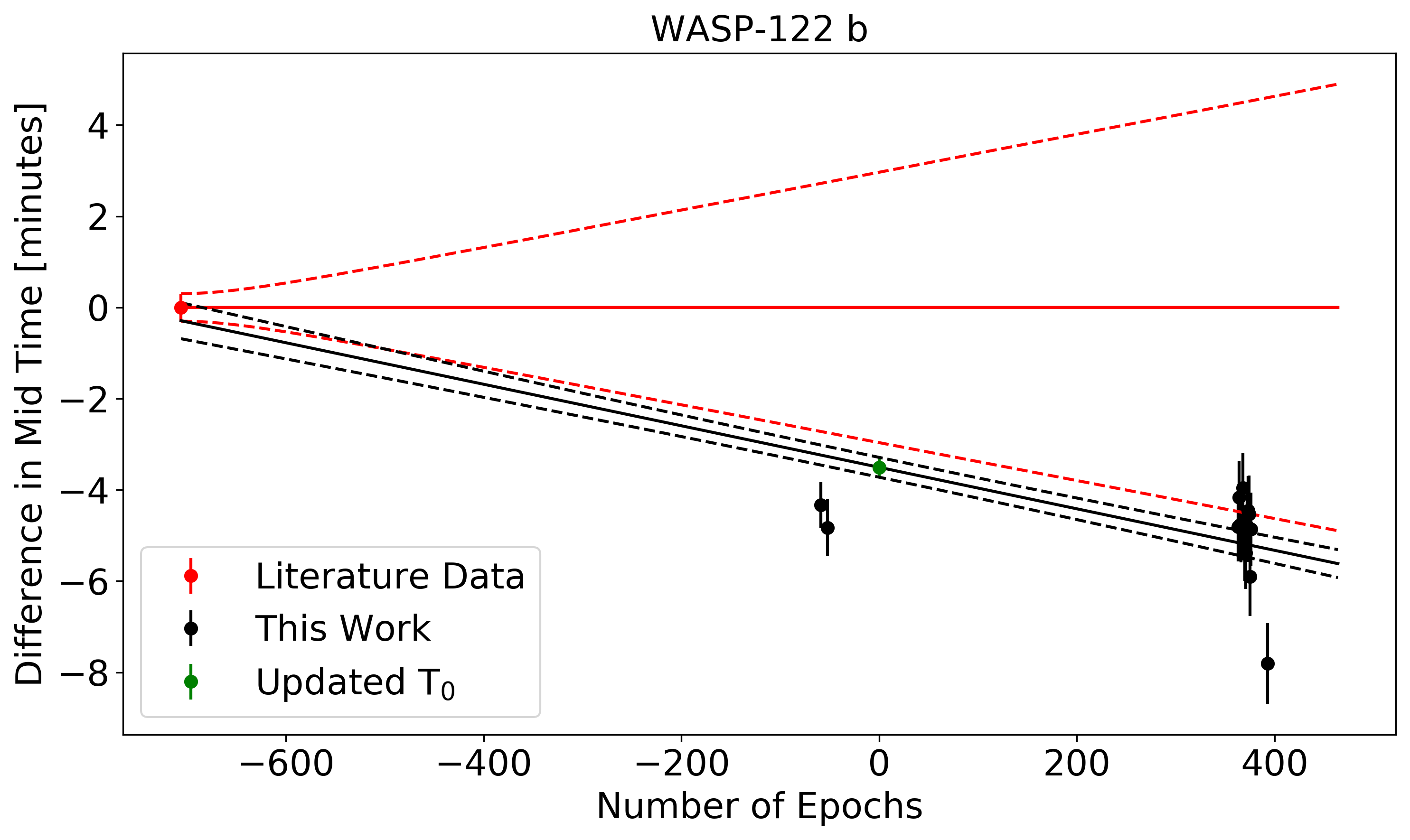}
    \caption{Observed minus calculated mid-transit times for all planet studied here. Transit midtime measurements from this work are shown in black, while literature T$_0$ values included in our calculation are in red. The green data point shows the updated T$_0$ reported. The black line denotes the new ephemeris of this work with the dashed lines showing the associated 1$\sigma$ uncertainties. For comparison, the previous literature ephemeris are given in red.}
    \label{oc_1}
\end{figure*}

Our analysis shows significant drifts in the transit times of all planets studied here, with only one planet (KPS-1 b) having observed transits within the 1 sigma errors on the expected time as shown in Figure \ref{oc_1}. Even in this case, the observed transit was considerably offset from that predicted. K2-237 b, KELT-15 b, WASP-45 b, WASP-83 b, WASP-119 b and WASP-122 b were observed by TESS in the first year of operations and we demonstrate the great potential the mission has for refining orbital parameters. The capability of TESS to provide accurate updated ephemeris for bright, short period planets has previously been shown in \cite{bouma}.

As the primary two year mission covers almost the entire sky, the ephemerides of many of the known planets could be updated once the data is released. For short period planets, TESS data gives multiple, high-precision, complete transits allowing the uncertainty on both the period and $T_0$ of these planets to be reduced. A summary of the findings for each planet is given below.

\textbf{CoRoT-6 b:} \cite{raetz} found that CoRoT planets seemed to have slightly underestimated uncertainties in their ephemerides and our analysis of CoRoT-6 b agrees with their findings. The last transit of CoRoT-6 b was found to be 23 minutes after the calculated transit time despite the predicted uncertainty being less than 3 minutes. The new observations help reduce the uncertainty on the transit time but we note there is not currently enough data to accurately constrain the period.

\textbf{K2-237 b:} There are two independent discovery papers for K2-237 b. Using the ephemeris data from \cite{soto} one could expect an uncertainty of 13 minutes in the transit mid time while \cite{smith} claims a greater precision on the period and thus predicts an uncertainty of 3 minutes. In reality we discover a shift of nearly 15 minutes and find our data to have a closer fit to the ephemeris of \cite{soto}. However, given the short period of the planet ($\sim$2.18 days), over 400 orbits have occurred since the discovery. The difference is therefore equivalent to an error in the period of $\sim$1.5 s, compared to a claimed uncertainty in the period of 0.5 s), and shows how slight errors in the accuracy of exoplanet ephemerides can lead to significant deviations from the expected transit time, highlighting the benefit of following-up targets on a regular basis.

\textbf{KELT-15 b:} This hot-Jupiter had not been re-observed with transit photometry since its discovery meaning the uncertainty in its transit time had risen to nearly 16 minutes. In the 4 years since, several hundred orbits had occurred and a 20 minute deviation from the expected transit time was found.

\textbf{KPS-1 b:} The newly observed transits for this planet were the only ones to fall within the 1 sigma errors in our sample. However, a deviation from the expected transit time of over 30 minutes was found, which is still a substantial residual. The uncertainty on the transit time of KPS-1 b is predicted to be less than 10 minutes until after the launch of Ariel in 2028.

\textbf{WASP-45 b:} The predicted uncertainty on WASP-45 b was around 5 minutes but had not been re-observed for several years. The O-C plot from ETD showed a slight divergence from that expected but it was not until the TESS data was analysed the full extent became clear with the transit arriving 15 minutes early.

\textbf{WASP-83 b:} For WASP-83 b, the newly observed transits occurred $\sim$30 minutes after the expected time, well outside the 1 sigma error of $\sim$12 minutes. Having not been observed since 2015, 300 orbits had passed. The literature orbital period differs by only 6 seconds from the updated value reported here, again highlighting the need for consistent follow-up.

\textbf{WASP-119 b:} With a reported discovery ephemeris in 2013 and an uncertainty of over 10 minutes, WASP-119 b was an obvious choice for follow-up. The combination of TESS data and a ground-based observations uncovered a drift of nearly 20 minutes over the 700 orbits since discovery.

\textbf{WASP-122 b:} Also known as KELT-14 b, this planet has two independent discovery papers \citep{turner,rodriguez}. These papers gave uncertainties of 3.4 and 4.9 minutes on the current transit mid time and the planet had not been re-observed since its discovery. Our observations found the transits of WASP-122 b to be occurring around 5 minutes early, just outside the 1 sigma errors. We  find our data has a better fit to the period from \cite{turner}.

Hence we detect significant variations in the observed transit time from the expected for most of the planets studied here. An overconfidence in the predicted transit time is a known issue and analysis of measured-to-predicted timing deviations of 21 exoplanets by \cite{mallonn} indicated a trend of slightly underestimated uncertainties in the ephemerides while \cite{raetz} made a similar finding for CoRoT planets. \cite{mallonn} found an average deviation of 1.4$\sigma$ while here we find a 2.2$\sigma$ divergence. We note that this is largely driven by CoRoT-6 b, which was observed to transit 7.7$\sigma$ from the expected time, and removing this planet reduces the average deviation to 1.5$\sigma$.

Here our analysis claims sub-second uncertainties on the periods of K2-237 b, WASP-45 b, WASP-83 b, WASP-119 b and WASP-122 b which should keep the uncertainty on the transit times of these planets to below 15 minutes until well after the launch of Ariel in 2028 (see Figure \ref{future_uncert}). Nevertheless, we would advocate further follow up of these planets, the others studied here and further planets with seemingly accurate ephemeris data to ensure errors are not underestimated. A cause of this can be the short baseline over which the period of the planet is determined when it is first discovered. When extrapolated over long times periods, even slight inaccuracies in the fitted period can cause significant deviations. Other sources of larger than expected uncertainties can be due to underestimated systematics in the data, stellar activity, tidal effects or transit timing variations (TTVs) due to other bodies in the system. Additionally, while the co-variance between T$_0$ and the period is minimised during fitting, it is non-zero. These effects can only be mitigated for by regularly observing transits over a long time period and to achieve this a well-organised ground and space-based campaign is required.

In any case, we emphasise that even in the event of the ephemerides being affected by astrophysical disturbances, our new ephemerides present the best available basis for future follow-up studies.

\section{Discussion}

The next generation of telescopes (JWST, Ariel, ELTs etc.) will require rigorous scheduling to minimise overheads and maximise science outputs. This means interesting science targets could see their observing priority degraded if their ephemerides are not accurate enough, even if they are excellent targets for atmospheric characterisation. Many currently known planets have large ephemeris uncertainties and analysis by \cite{dragomir_tess} suggests many TESS targets will have errors of \textgreater 30 minutes less than a year after discovery due to the short baseline of TESS observations. To maintain, and verify, the ephemeris of these planets will require follow-up observations over the coming years from the ground and from space. 

Given the brightness of the host stars and the large transit depths of many of these planets, modestly sized telescopes can play a crucial role in this activity. To achieve regular observations of all known exoplanets, telescope time must be efficiently utilised. As it is not always clear when the most recent observations of a target occurred, follow-up observations need to be coordinated to maximise the effectiveness of the data. Community-wide citizen science efforts, such as the Exoplanet Transit Database and Exoplanet Timing Survey \citep{zellem} and ExoClock, need to be established to create a network of individuals and groups to monitor new discoveries. This need not be limited to citizen astronomers but should include all research groups with access to telescopes of all sizes. This organised and structured approach should build upon the work of existing schemes.

Researcher engagement with citizen astronomers, citizen science and in educational outreach offers an excellent opportunity to support future space missions. Stimulating this engagement and devising a coordinated approach to maintaining exoplanet ephemeris will be imperative in the coming years. Projects such as the ETS, ExoClock and ORBYTS need to become more widespread and methodical in their approach to transit follow-up (the selection of targets here was somewhat ad-hoc and this is what needs to be avoided). Finally, it is critical that data from these observations is publicly available, especially for planets which display deviations in transit mid-time. 

Here we homogeneously analyse the data but the reduction method has differed between observers which can lead to inconsistencies. The ETD provides a rating of the quality of the uploaded, from 1-5, but its vetting is perhaps not as extensive as other databases such as the Minor Planet Center. A systematic approach is required, with guidelines that ensure all data is processed in the correct manner. For exoplanet observations, the choice of comparison stars, the provision of correct timing and overall consistency are paramount. Performing such quality checks on the data can be complex, requiring significant data storage and processing capabilities, but are critical if high precision ephemerides are to be obtained. Thus future projects should allow the submission of raw images, along with the necessary calibration files, to allow for the data to homogeneously reduced and analysed and, if an observer wishes to download data, the output format needs to be consistent to ensure efficacy. Being able to accept, and return, various data products from the raw frames to the light curve will increase the functionality of such a project. Alternatively, easy to use codes which automatically process the data without the need for human input (e.g. in the choice of comparison stars) could be used. ExoClock is expected to be expanded to provide such a platform for planets that could potentially be studied by Ariel.

\vspace{-5mm}
\section{Conclusions}

We present follow-up observations of eight exoplanets with large uncertainties in their predicted transit time via a network of robotic ground-based telescopes and data from TESS. We refine the ephemeris data for these planets and for seven of them, find that the observed transit time was outside the predicted uncertainties. This can only be mitigated for by regularly following up targets and, given the number of planets that are expected to be detected in the coming years, such an effort will require a large amount of telescope time. Therefore a coordinated approach is required and citizen astronomers and educational outreach provide an excellent opportunity to contribute towards this effort. Schemes which stimulate this engagement will be crucial in maintaining transit times for the next generation of telescopes.
\vspace{-5mm}
\section{Acknowledgements}

The authors wish to thank Dr Ehsan Pedram from Preston Manor School and Martin Yates from Beal High School for their dedication in organising the outreach sessions, devoting their spare time for the benefit of their students. We also thank Telescope Live for kindly proving access to their telescope network, without which this project would not have been possible. Access to the LCO network was provided to the ORBYTS programme by the Faulkes Telescope Project which is coordinated by Cardiff University and Swansea University. Additionally, we thank Paul Edwards for supplying data storage solutions. This paper includes data collected by the TESS mission which is funded by the NASA Explorer Program. TESS data is publicly available via the Mikulski Archive for Space Telescopes (MAST). This research has also used the NASA Exoplanet Archive, which is operated by the California Institute of Technology, under contract with the National Aeronautics and Space Administration through the Exoplanet Exploration Program. This work has been funded through the European Union's Horizon 2020 research and innovation programme (grant agreement No 758892, ExoAI), and with the STFC grants ST/P000282/1, ST/P002153/1, ST/S002634/1 and ST/T001836/1. Finally, we thank the MAPS Faculty at UCL for their partial funding of the ORBYTS programme.

\vspace{-5mm}
\section{Affiliations}
$^{1}$Department of Physics and Astronomy, University College London, Gower Street, London, WC1E 6BT, UK\\
$^{2}$Department of Physics (Atmospheric, Oceanic and Planetary Physics), University of Oxford, Parks Rd, Oxford, OX1 3PU, UK\\
$^{3}$Preston Manor High School, Carlton Avenue East, Wembley, HA9 8NA, UK\\
$^{4}$Beal High School, Woodford Bridge Road, Ilford, Essex, IG4 5LP, UK\\
$^{5}$Royal Observatory Greenwich, London, UK\\
$^{6}$Birkbeck, University of London, Malet Street, London, WC1E 7HX, UK\\
$^{7}$Spaceflux Ltd., 71-75 Shelton Street, Covent Garden, London, WC2H 9JQ, UK\\
$^{8}$Leibniz-Institut f\"ur Astrophysik Potsdam, An der Sternwarte 16, D-14482 Potsdam, Germany\\
$^{9}$Club d'Astronomie de Wittelsheim, Wittelsheim, France\\
$^{10}$Santa Maria de Montmagastrell Remote Observatory\\
$^{11}$Observatory of Baronnies Proven\c{c}ales, 05150 Moydans, France\\
$^{12}$Observatoire Sadr Chili\\
$^{13}$Les Barres Observatory, Lamanon, France\\
$^{14}$Facultad de Ciencias Astron\'omicas y Geof\'isicas, Universidad Nacional de La Plata, Paseo del Bosque S/N-1900 La Plata, Argentina\\
$^{15}$Instituto de Astrof\'isica de La Plata (CCT La Plata - CONICET/UNLP), Argentina\\
$^{16}$Rarotonga Observatory, Cook Islands\\
$^{17}$El Sauce Observatory, Coquimbo Province, Chile\\
$^{18}$Observatori Astron\'omic Albany\'a, Cam\'i de Bassegoda, Albany\'a 17733 , Girona, Spain\\
$^{19}$Observatory Ca l'Ou, San Mart\'i Sesgueioles\\
$^{20}$National Youth Space Center, Goheung, Jeollanam-do, 59567, Republic of Korea,\\
$^{21}$OPERA Observatory, 33820 Saint Palais, France,\\
$^{22}$Observatory of Josef Sadil,  Havl\'i\v{c}kova 514, Sedl\v{c}any  CZ-264 01, Czech Republic\\
$^{23}$Observatorio de Elche, Elche, Spain\\
$^{24}$Taurus Hill Observatory, 79480 Kangaslampi, Finland\\
$^{25}$Observatoire de Vaison la Romaine, Au Palis, 84110 Vaison la Romaine, France\\
$^{26}$Anunaki Observatory, Manzanares El Real, Spain\\
$^{27}$Montcabrer Observatory\\
$^{28}$Raetz Observatory, Stiller Berg 6, 98587 Herges-Hallenberg, Germany\\
$^{29}$International Citizen Observatory, Germany\\
$^{30}$British Astronomical Association, Burlington House, Piccadilly, London W1J 0DU, UK\\
$^{31}$Variable Star and Exoplanet Section of Czech Astronomical Society, Vset\'insk\'a 941/78, 757 01 Vala\v{s}sk\'e Mezi\v{r}\'i\v{c}\'i, Czech Republic\\
$^{32}$Blue Skies Space Ltd., 69 Wilson Street, London, EC2A 2BB, UK\\
$^{33}$University College London Observatory, Mill Hill, London, NW7 2QS, UK\\


\clearpage
\onecolumn
\setlength\LTleft{0pt}
\setlength\LTright{0pt}
\tiny
\begin{longtable}{@{\extracolsep{\fill}}ccccccccc@{}}
    \caption{General information about the observations conducted and analysed in this work. The label corresponds to that given in Figures \ref{WASP-122}-\ref{WASP-45}.}
    \label{tab:all observations}\\\hline
         Planet & Date & Telescope & Filter & Exposure Time [s] & Mid Time [BJD] & Mid Time Error & Epoch & Label\\\hline
         & 09 July 2010 & ETD, Sauer & Clear & 90 & 2455386.52167 & 0.00085 & 63 & G1\\
          & 17 July 2010 & ETD, Sauer & Clear & 90 & 2455395.40632 & 0.00074 & 64 & G2\\
         CoRoT-6 b & 17 June 2015 & ETD, Molina & Clear & 300 & 2457190.50697 & 0.00698 & 266 & G3\\
          & 19 July 2018 & ETD, Kang & R & 300 & 2458319.10958 & 0.00102 & 393 & G4\\
        & 22 June 2019 & ETD, Evans &  Clear & 300 & 2458656.80291 & 0.00242 & 431 & G5\\\hline
        
         K2-237 b & 16 April & El Sauce & V & 30 & 2458589.73380 & 0.00061 & 370 & G1 \\ 
         & &\multirow{7}{*}{TESS S12} & \multirow{7}{*}{I} & \multirow{7}{*}{120} & 2458631.16350 & 0.00095 & 389 & T1\\
         &  &  & &  & 2458633.34306 & 0.00077 & 390 & T2\\
         & 26 May 2019  &  & & & 2458635.52308 & 0.00078 & 391 & T3\\
         &  till  &  & &  & 2458637.70413 & 0.00095 & 392 & T4\\
         &  18 June 2019&  & &  & 2458646.42401 & 0.00089 & 396 & T5\\
         &  &  & &  & 2458648.60616 & 0.00097 & 397 & T6\\
         &  &  & &  & 2458650.78837 & 0.00090 & 398 & T7\\\hline
         
          & &\multirow{12}{*}{TESS S7,9} & \multirow{12}{*}{I} & \multirow{12}{*}{120} & 2458494.13536 & 0.00068 & -10 & T1 \\
                    &  &  & &  & 2458497.46521 & 0.00061 & -9 & T2\\
                    &  &  & &  & 2458500.79344 & 0.00062 & -8 & T3\\
                    &  &  & &  & 2458507.45364 & 0.00067 & -6 & T4\\
                    & 08 January 2019 &  & &  & 2458510.78149 & 0.00068 & -5 & T5\\
         KELT-15 b & till &  & &  & 2458514.11222 & 0.00066 & -4 & T6\\
                    & 27 March 2019 &  & &  & 2458547.40593 & 0.00073 & 6 & T7\\
                    &  &  & &  & 2458550.73620 & 0.00077 & 7 & T8\\
                    &  &  & &  & 2458554.06598 & 0.00072 & 8 & T9\\
                    &  &  & &  & 2458560.72358 & 0.00078 & 10 & T10\\
                    &  &  & &  & 2458564.05458 & 0.00074 & 11 & T11\\
                    &  &  & &  & 2458567.38262 & 0.00075 & 12 & T12\\
                   & 19 February 2019 & Warrumbungle & V & 120 & 2458534.08713 & 0.0024 & 2 & G1\\
                    & 21 March 2019 & Warrumbungle & V & 120 & 2458564.05478 & 0.00358 & 11 & G2\\\hline
                    
         &	12 March 2019	&	ETD,	Jongen	&	Clear & 120 & 2458554.34979 & 0.00064 & -10 & G1 \\
&	22 March 2019	&	ETD,	Wunsche	&	V & 120 & 2458564.58226 & 0.00096 & -4 & G2\\
&	22 March 2019	&	ETD,	Wunsche	&	Clear	& 120 & 2458564.58759 & 0.00164 & -4 & G3\\
&	22 March 2019	&	ETD,	Raetz	&	Clear	& 60 & 2458564.58547 & 0.00090 & -4 & G4\\
&	22 March 2019	&	ETD,	Jongen	&	Clear	& 120 & 2458564.58697 & 0.00093 & -4 & G5\\
&	22 March 2019	&	ETD,	Guerra	&	CBB	& 120 & 2458564.58415 & 0.00060 & -4 & G6\\
&	29 March 2019	&	ETD,	Wunsche	&	Clear	& 120 & 2458571.40954 & 0.00182 & 0 & G7\\
&	29 March 2019	&	ETD,	Jongen	&	Clear	& 120 & 2458571.41164 & 0.00047 & 0 & G8\\
&	29 March 2019	&	ETD,	Friedli/Kropf &	V & 60 & 2458571.40894 & 0.00025 & 0 & G9\\
&	29 March 2019	&	ETD,	Watkins	&	R & 30 & 2458571.41121 & 0.00065 & 0 & G10	\\
KPS-1 b &	29 March 2019	&	ETD,	Guerra	&	I & 180 & 2458571.41002 & 0.00098 & 0 & G11	\\
&	20 April 2019	&	ETD,	Wunsche	&	Clear & 120 & 2458593.59136 &  0.00147 & 13 & G12\\
&	20 April 2019	&	ETD,	Jongen	&	Clear & 120 & 2458593.59541 & 0.00113 & 13 & G13\\
&	02 May 2019	&	ETD,	Raetz	&	Clear & 120 & 2458605.53684 & 0.00099 & 20 & G14\\
&	14 May 2019	&	ETD,	Bretton	&	I & 120 & 2458617.47815 & 0.00085 & 27 & G15\\
&	14 May 2019	&	ETD,	Guerra	&	V & 180 & 2458617.48126 & 0.00066 & 27 & G16	\\
&	14 May 2019	&	ETD,	Raetz	&	Clear & 120 & 2458617.48309 & 0.00079 & 27 & G17\\
&	14 May 2019	&	ETD,	Bosch	&	V & 200 & 2458617.48291 & 0.00035 & 27 & G18 \\
&	14 May 2019	&	ETD,	Watkins	&	V & 30 & 2458617.48255 & 0.00162 & 27 & G19 \\
&   31 May 2019 & ETD, Bretton & Clear & 120 & 2458634.54617 & 0.00026 & 37 & G20 \\
&   31 May 2019 & ETD, Jongen & Clear & 120 & 2458634.54662 & 0.00050 & 37 & G21 \\\hline

    & 15 August 2011 & ETD, Evans & Clear & 60 & 2455782.01348 & 0.00097 & -550 & G1\\
    & 16 July 2012 & ETD, Sauer & R & 60 & 2456119.62831 & 0.00097 & -442 & G2\\ 
    & 27 December 2016 & ETD, Lajus & R & 10 & 2457660.78567 & 0.00056 & 51 &G3\\ 
    & &\multirow{8}{*}{TESS S2} & \multirow{8}{*}{I} & \multirow{8}{*}{120} & 2458354.77330 & 0.00093 & 273 & T1 \\
    WASP-45 b  & & & & & 2458357.89881 & 0.00085 & 274 & T2\\ 
                    & & & & & 2458361.02613 & 0.00079 & 275 & T3\\ 
                    & 23 August 2018 & & & & 2458364.15241 & 0.00110 & 276 & T4\\ 
                    & till & & & & 2458370.40257 & 0.00089 & 278 & T5\\ 
                    & 20 September 2019& & & & 2458373.52974 & 0.00096 & 279 & T6\\ 
                    & & & & & 2458376.65588 & 0.00077 & 280 & T7\\ 
                    & & & & & 2458379.78198 & 0.00088 & 281 & T8\\
                    \hline
                    
      & 02 April 2019 & El Sauce & R & 120 & 2458205.73765 & 0.00083 & 213 & G1\\
                    & 28 March 2019 &\multirow{4}{*}{TESS S10} & \multirow{4}{*}{I} & \multirow{4}{*}{120} & 2458573.61364 & 0.00090 & 287 & T1 \\
      WASP-83 b  & till & & & & 2458578.58484 & 0.00095 & 288 & T2\\
                    & 22 April 2019 & & & & 2458588.52739 & 0.00114 & 290 & T3\\
                    &  & & & & 2458593.49776 & 0.00094 & 291 & T4\\ \hline
                    
    & & \multirow{54}{*}{TESS S1-4,7,11} & \multirow{54}{*}{I} &  \multirow{54}{*}{120} & 2458327.40896 & 0.00064 & -33 & T1\\
                    & & & & & 2458329.90921 & 0.00063 & -32 & T2\\
                    & & & & & 2458332.40913 & 0.00069 & -31 & T3\\
                    & & & & & 2458334.90914 & 0.00059 & -30 & T4\\
                    & & & & & 2458337.40888 & 0.00071 & -29 & T5\\
                    & & & & & 2458342.40948 & 0.00063 & -27 & T6\\
                    & & & & & 2458344.90830 & 0.00062 & -26 & T7\\
                    & & & & & 2458347.40851 & 0.00147 & -25 & T8\\
                    & & & & & 2458349.90839 & 0.00063 & -24 & T9\\
                    & & & & & 2458352.40769 & 0.00061 & -23 & T10\\
                    & & & & & 2458354.90856 & 0.00080 & -22 & T11\\
                    & & & & & 2458357.40699 & 0.00074 & -21 & T12\\
                    & & & & & 2458359.90699 & 0.00071 & -20 & T13\\
                    & & & & & 2458362.40801 & 0.00070 & -19 & T14\\
                    & & & & & 2458364.90547 & 0.00082 & -18 & T15\\
                    & & & & & 2458369.90681 & 0.00067 & -16 & T16\\
                    & & & & & 2458372.40655 & 0.00079 & -15 & T17\\
                    & & & & & 2458374.90585 & 0.00074 & -14 & T18\\
                    & & & & & 2458377.40605 & 0.00064 & -13 & T19\\
                    & & & & & 2458379.90440 & 0.00071 & -12 & T20\\
                    & & & & & 2458387.40490 & 0.00071 & -9 & T21\\
                    & & & & & 2458389.90515 & 0.00068 & -8 & T22\\
                    & & & & & 2458392.40410 & 0.00066 & -7 & T23\\
                    & & & & & 2458394.90473 & 0.00087 & -6 & T24\\
                    & & & & & 2458397.40311 & 0.00064 & -5 & T25\\
                    & 27 July 2018 & & & & 2458399.90351 & 0.00064 & -4 & T26\\
     WASP-119 b     & till & & & & 2458402.40275 & 0.00058 & -3 & T27\\
                    & 18 May 2019 & & & & 2458404.90377 & 0.00059 & -2 & T28\\
                    & & & & & 2458412.40318 & 0.00071 & 1 & T29\\
                    & & & & & 2458414.90327 & 0.00070 & 2 & T30\\
                    & & & & & 2458417.40304 & 0.00064 & 3 & T31\\
                    & & & & & 2458422.40234 & 0.00068 & 4 & T32\\
                    & & & & & 2458424.90216 & 0.00060 & 6 & T33\\
                    & & & & & 2458427.40244 & 0.00066 & 7 & T34\\
                    & & & & & 2458429.90269 & 0.00063 & 8 & T35\\
                    & & & & & 2458432.40216 & 0.00062 & 9 & T36\\
                    & & & & & 2458434.90223 & 0.00069 & 10 & T37\\
                    & & & & & 2458492.39645 & 0.00117 & 33 & T38\\
                    & & & & & 2458494.89668 & 0.00129 & 34 & T39\\
                    & & & & & 2458497.39677 & 0.00123 & 35 & T40\\
                    & & & & & 2458499.89584 & 0.00128 & 36 & T41\\
                    & & & & & 2458502.39712 & 0.00112 & 37 & T42\\
                    & & & & & 2458504.89546 & 0.00127 & 38 & T43\\
                    & & & & & 2458507.39514 & 0.00130 & 39 & T44\\
                    & & & & & 2458509.89502 & 0.00123 & 40 & T45\\
                    & & & & & 2458512.39476 & 0.00122 & 41 & T46\\
                    & & & & & 2458514.89459 & 0.00130 & 42 & T47\\
                    & & & & & 2458602.38802 & 0.00085 & 77 & T48\\
                    & & & & & 2458604.88824 & 0.00088 & 78 & T49\\
                    & & & & & 2458607.38852 & 0.00076 & 79 & T50\\
                    & & & & & 2458614.88619 & 0.00089 & 82 & T51\\
                    & & & & & 2458617.38771 & 0.00084 & 83 & T52\\
                    & & & & & 2458619.88570 & 0.00083 & 84 & T53\\
                    & & & & & 2458622.38559 & 0.00076 & 85 & T54\\
                    & 07 March 2019 & Warrumbungle & V & 60 & 2458549.89284 & 0.00596 & 56 & G1\\\hline
                    
        &  18 January 2017 & ETD, Evans & R & 75 & 2457771.62839 & 0.00035 & -59 & G1\\
                    &  30 January 2017 & ETD, Evans & R & 80 & 2457783.59845 & 0.00044 & -52 & G2\\
                    &  02 March 2019 & LCO & V & 5 & 2458544.57156 & 0.00061 & 393 & G3\\ 
                    
                    & &\multirow{14}{*}{TESS S7} & \multirow{14}{*}{I} & \multirow{14}{*}{120} & 2458493.27195 & 0.00052 & 363 & T1\\
                    & & & & & 2458494.98245 & 0.00056 & 364 & T2\\
                    & & & & & 2458496.69209 & 0.00056 & 365 & T3\\
                    & & & & & 2458498.40213 & 0.00055 & 366 & T4\\
                    & 08 January 2019 & & & & 2458500.11221 & 0.00056 & 367 & T5\\
         WASP-122 b & till & & & & 2458501.82282 & 0.00054 & 368 & T6\\
                    & 01 February 2019 & & & & 2458505.24207 & 0.00055 & 370 & T7\\
                    & & & & & 2458506.95200 & 0.00055 & 371 & T8\\
                    & & & & & 2458508.66246 & 0.00054 & 372 & T9\\
                    & & & & & 2458510.37276 & 0.00053 & 373 & T10\\
                    & & & & & 2458512.08277 & 0.00058 & 374 & T11\\
                    & & & & & 2458513.79187 & 0.00060 & 375 & T12\\
                    & & & & & 2458515.50265 & 0.00056 & 376 & T13\\ \hline
\end{longtable}

\normalsize
\twocolumn

\begin{table*}
    \centering
    \begin{tabular}{lllll} \hline
     Parameter & Units & CoRoT-6 b & K2-237 b & KELT-15 b \\ \hline
R$_{*}$ & Star Radius [R$_{\odot}$] & 1.025$\pm0.026 ^{\dagger}$ & 1.43$^{+0.06}_{-0.07}$ $^{\ddagger}$ & 1.481$^{+0.091}_{-0.041}$ $^{\star}$ \\
M$_{*}$ & Star Mass [M$_{\odot}$] & 1.05$\pm$0.05  & 1.28$^{+0.03}_{-0.04}$ $^{\ddagger}$ & 1.181$^{+0.051}_{-0.050}$ $^{\star}$ \\
T$_{eff}$ & Star Effective Temperature [K] & 6090$\pm$70$ ^{\dagger}$  & 6257$\pm100 ^{\ddagger}$ & 6003$^{+56}_{-52}$ $^{\star}$ \\
Fe/H & Star Metallicity & -0.2$\pm$0.1$^{\dagger}$  & 0.14 $\pm0.05 ^{\ddagger}$ & 0.047$\pm0.032 ^{\star}$ \\
log(g$_{*}$) & Star Surface Gravity [cgs] & 4.43$\pm0.1 ^{\dagger}$ & 4.24$\pm0.1$ $^{\ddagger}$ & 4.168$^{+0.019}_{-0.044}$ $^{\star}$ \\
$\rho_{*}$ & Star Density [$\rho_{\odot}$] & 1.31$\pm0.09 ^{\dagger}$ & 0.144$^{+0.017}_{-0.014}$ $^{\ddagger}$ & 0.514$^{+0.034}_{-0.076}$ $^{\star}$ \\ \hline
R$_{P}$ & Planet Radius [R$_{J}$] & 1.166$\pm0.035$ $^{\dagger}$ & 1.6944$^{+0.0118}_{-0.0103}$ & 1.4745$^{+0.0033}_{-0.0415}$ \\
R$_{P}$ &  Planet Radius [R$_{\oplus}$] & 13.068$\pm0.392^{\dagger}$ & 18.5935$^{+0.1298}_{-0.1136}$ & 16.18020$^{+0.00364}_{-0.45555}$ \\
R$_{P}$/R$_{*}$ & Planet Radius in Stellar Radii & 0.1198$\pm0.0036$ $^{\dagger}$ & 0.12342$^{+0.00086}_{-0.00075}$ & 0.1000 $^{+0.00022}_{-0.00281}$\\
$\delta$ & Transit Depth & 0.17$\pm0.0008 ^{\dagger}$ & 0.01523$^{+0.00021}_{-0.00019}$ & 0.0100$^{+0.00005}_{-0.00056}$ \\
M$_{P}$ & Planet Mass [M$_{J}$] & 2.96$\pm0.34^{\dagger}$ & 1.6$\pm0.11 ^{\ddagger}$ & 0.910 $^{+0.210}_{-0.220}$ $^{\star}$ \\
M$_{P}$ & Planet Mass [M$_{\oplus}$] & 940.74$\pm$108.06$ ^{\dagger}$ & 509$\pm35 ^{\ddagger}$ & 289$^{+67}_{-70}$ $^{\star}$ \\

P & Period [days] & 8.886621$\pm$0.0000063 & 2.1805358$\pm$0.0000010 & 3.329468$\pm$0.000012 \\
T$_0$ & Transit Mid Time [BJD$_{TDB}$] & 2454826.66625$\pm$0.00054 & 2457782.93422$\pm$0.00014 & 2458527.42971$\pm$0.00017 \\

a & Semi-major axis [AU] & 0.0855$\pm0.0015$ $ ^{\dagger}$ & 0.03602$^{+0.00023}_{-0.00042}$ & 0.04778$^{+0.00022}_{-0.0082}$ \\
a/R$_{*}$ & Semi-major axis in stellar radii & 17.94$\pm0.33 ^{\dagger}$ & 5.6138$^{+0.0364}_{-0.0655}$ & 6.940$^{+0.032}_{-0.118}$\\
i & Inclination [degrees] & 89.07$\pm0.3 ^{\dagger}$ & 84.888$^{+0.024}_{-0.264}$ & 88.339 $^{+0.169}_{-0.052}$ \\
b & Impact Parameter & 0.291$\pm0.091 ^{\dagger}$ & 0.500$\pm0.00583$ & 0.1946$\pm$0.0048 \\
e & Eccentricity & Fixed to zero & Fixed to zero & Fixed to zero \\\hline
\multicolumn{5}{l}{$^{\dagger}$\cite{fridlund}, $^{\ddagger}$\cite{soto}, $^{\star}$\cite{rodriguez}}\\\hline
    \end{tabular}
    \caption{Summary of updated system parameters for CoRoT-6 b, K2-237 b and KELT-15 b.}
    \label{tab:para 1}
\end{table*}

\begin{table*}
    \centering
    \begin{tabular}{lllll} \hline
     Parameter & Units & KPS-1 b & WASP-45 b & WASP-83 b \\\hline
R$_{*}$ & Star Radius [R$_{\odot}$] & 0.907 $^{+0.086}_{-0.082}$ $^{\dagger}$ & 0.945 $\pm0.087$ $^{\dagger}$ & 1.05$^{+0.06}_{-0.04}$ $^{\ast}$ \\
M$_{*}$ & Star Mass [M$_{\odot}$] & 0.892 $^{+0.090}_{-0.100}$ $^{\ddagger}$ & 0.909$\pm$0.060 $^{\ddagger}$ & 1.11$\pm$0.09 $^{\ast}$ \\
T$_{eff}$ & Star Effective Temperature [K] & 5165 $\pm90$ $^{\dagger}$ & 5140$\pm$200 $^{\ddagger}$ & 5510$\pm$110 $^{\ast}$ \\
Fe/H & Star Metallicity & 0.22 $\pm$0.13 $^{\dagger}$ & 0.43 $\pm$0.06 $^{\star}$ & 0.29$\pm$0.12 $^{\ast}$ \\
log(g$_{*}$) & Star Surface Gravity [cgs] & 4.47 $\pm$0.06 $^{\dagger}$ & 4.43$\pm$0.18 $^{\star}$ & 4.44 $^{+0.02}_{-0.04}$ $^{\ast}$ \\
$\rho_{*}$ & Star Density [$\rho_{\odot}$] & 1.68 $^{+0.41}_{-0.32}$ $^{\dagger}$ & 1.08$\pm$0.25 $^{\star}$ & 1.40 $^{+0.10}_{-0.18}$ $^{\ast}$ \\ \hline
R$_{P}$ & Planet Radius [R$_{J}$] & 1.03 $^{+0.13}_{-0.12}$ $^{\dagger}$ & 1.079 $^{+0.047}_{-0.016}$ & 1.039 $^{+0.012}_{-0.008}$ \\
R$_{P}$ & Planet Radius [R$_{E}$] & 11.5 $^{+1.5}_{-1.3}$ $^{\dagger}$ & 11.849$^{+0.519}_{-0.173}$ & 11.402 $^{+0.129}_{-0.086}$ \\
R$_{P}$/R$_{*}$ & Planet Radius in Stellar Radii & 0.1143 $^{+0.0037}_{-0.0034}$ $^{\dagger}$ & 0.1149$^{+0.0050}_{-0.0017}$ & 0.09947$^{+0.00113}_{-0.00075}$ \\
$\delta$ & Transit Depth & 0.01306 $^{+0.00165}_{-0.00170}$ $^{\dagger}$ & 0.01319 $^{+0.00116}_{-0.00039}$ & 0.009895 $^{+0.00224}_{-0.00015}$\\
M$_{P}$ & Planet Mass [M$_{J}$] & 1.090 $^{+0.086}_{-0.087}$ $^{\dagger}$ & 1.007$\pm$0.053 $^{\ddagger}$ & 0.30$\pm$0.03 $^{\ast}$\\
M$_{P}$ & Planet Mass [M$_{\oplus}$] & 346.4 $^{+27.3}_{-27.7}$ $^{\dagger}$ & 320.0421$\pm$16.844 $^{\ddagger}$ & 95$\pm$10 $^{\ast}$\\

P & Period [days] & 1.7063270$\pm$0.0000036 & 3.12607637$\pm$0.00000060 & 4.97129175$\pm$0.00000051\\
T$_0$ & Transit Mid Time [BJD$_{TDB}$] & 2458571.41092$\pm$0.00035 & 2457501.35468$\pm$0.00022 & 2457146.85256$\pm$0.00013 \\

a & Semi-major axis [AU] & 0.0269 $\pm$ 0.001 $^{\dagger}$ & 0.04295 $^{+0.00048}_{-0.00161}$ & 0.0592 $^{+0.0019}_{-0.0009}$ \\
a/R$_{*}$ & Semi-major axis in stellar radii & 6.38$^{+0.77}_{-0.72}$ $^{\dagger}$ & 9.78$^{+0.11}_{-0.36}$ & 12.1 3$^{+0.39}_{-0.19}$ \\
i & Inclination [degrees] & 83.20$^{+0.88}_{-0.90}$ $^{\dagger}$ & 84.98$^{+0.03}_{-0.31}$ & 88.91$^{+0.50}_{-0.43}$ \\
b & Impact Parameter & 0.754 $\pm0.049$ $^{\dagger}$ & 0.855$\pm$0.032 & 0.28 $\pm0.007$ \\
e & Eccentricity & Fixed to zero & Fixed to zero & Fixed to zero \\ \hline
\multicolumn{5}{l}{$^{\dagger}$\cite{burdanov}, $^{\ddagger}$\cite{anderson}, $^{\star}$\cite{mortier}, $^{\ast}$\cite{hellier}}\\\hline
    \end{tabular}
    \caption{Summary of updated system parameters for KPS-1 b, WASP-45 b and WASP-83 b.}
    \label{tab:para 2}
\end{table*}

\begin{table*}
    \centering
    \begin{tabular}{llll} \hline
     Parameter & Units & WASP-119 b & WASP-122 b \\\hline
R$_{*}$ & Star Radius [R$_{\odot}$] & 1.2$\pm$0.1 $^{\dagger}$ & 1.52$\pm$0.03 $^{\ddagger}$\\
M$_{*}$ & Star Mass [M$_{\odot}$] & 1.02$\pm$ 0.06 $^{\dagger}$ & 1.239$\pm$0.039 $^{\ddagger}$ \\
T$_{eff}$ & Star Effective Temperature [K] & 5650$\pm$ 100 $^{\dagger}$ & 5720$\pm$130 $^{\ddagger}$ \\
Fe/H & Star Metallicity & 0.14$\pm$0.10 $^{\dagger}$ & 0.32$\pm$ $^{\ddagger}$ \\
log(g$_{*}$) & Star Surface Gravity [cgs] & 4.26$\pm$0.08 $^{\dagger}$ & 4.166$\pm$0.016 $^{\ddagger}$ \\
$\rho_{*}$ & Star Density [$\rho_{\odot}$] & 0.76 $\pm$ 0.25 $^{\dagger}$ & 0.495$\pm$0.025 $^{\ddagger}$ \\ \hline
R$_{P}$ & Planet Radius [R$_{J}$] & 1.3542 $^{+0.0039}_{-0.0023}$ & 1.712 $^{+0.009}_{-0.006}$ \\
R$_{P}$ & Planet Radius [R$_{E}$] & 14.860 $^{+0.043}_{-0.0255}$ & 18.785 $^{+0.102}_{-0.068}$ \\
R$_{P}$/R$_{*}$ & Planet Radius in Stellar Radii & 0.11344 $^{+0.00032}_{-0.00019}$
& 0.1132 $^{+0.0006}_{-0.0004}$ \\
$\delta$ & Transit Depth & 0.012868 $^{+0.000074}_{-0.000044}$ & 0.012815 $^{+0.000139}_{-0.000093}$ \\
M$_{P}$ & Planet Mass [M$_{J}$] & 1.23$\pm$0.08 $^{\dagger}$ & 1.284$\pm$0.032 $^{\ddagger}$\\
M$_{P}$ & Planet Mass [M$_{\oplus}$] & 391$\pm$25 $^{\dagger}$ & 408.1$\pm$10.2 $^{\ddagger}$ \\

P & Period [days] & 2.4998052 $\pm$0.0000017 & 1.71005344$\pm$0.00000032 \\
T$_0$ & Transit Mid Time [BJD$_{TDB}$] & 2458409.903247$\pm$0.000086 & 2457872.52231$\pm$0.00015\\

a & Semi-major axis [AU] & 0.03851 $^{+0.00008}_{-0.00021}$ & 0.03024 $^{+0.00007}_{-0.00019} $\\
a/R$_{*}$ & Semi-major axis in stellar radii & 6.902 $^{+0.014}_{-0.037}$ & 4.279$^{+0.01}_{-0.027}$ \\
i & Inclination [degrees] & 86.99 $^{+0.01}_{-0.11}$ &  78.595 $^{+0.009}_{-0.010}$\\
b & Impact Parameter & 0.362 $\pm0.002$ & 0.846 $\pm0.005$ \\
e & Eccentricity & Fixed to zero &  Fixed to zero\\ \hline
\multicolumn{4}{l}{$^{\dagger}$\cite{maxted}, $^{\ddagger}$\cite{turner}}\\\hline
    \end{tabular}
    \caption{Summary of updated system parameters for WASP-119 b and WASP-122 b.}
    \label{tab:para 3}
\end{table*}

\clearpage

\begin{figure}
    \centering
    \includegraphics[width = 0.47\textwidth]{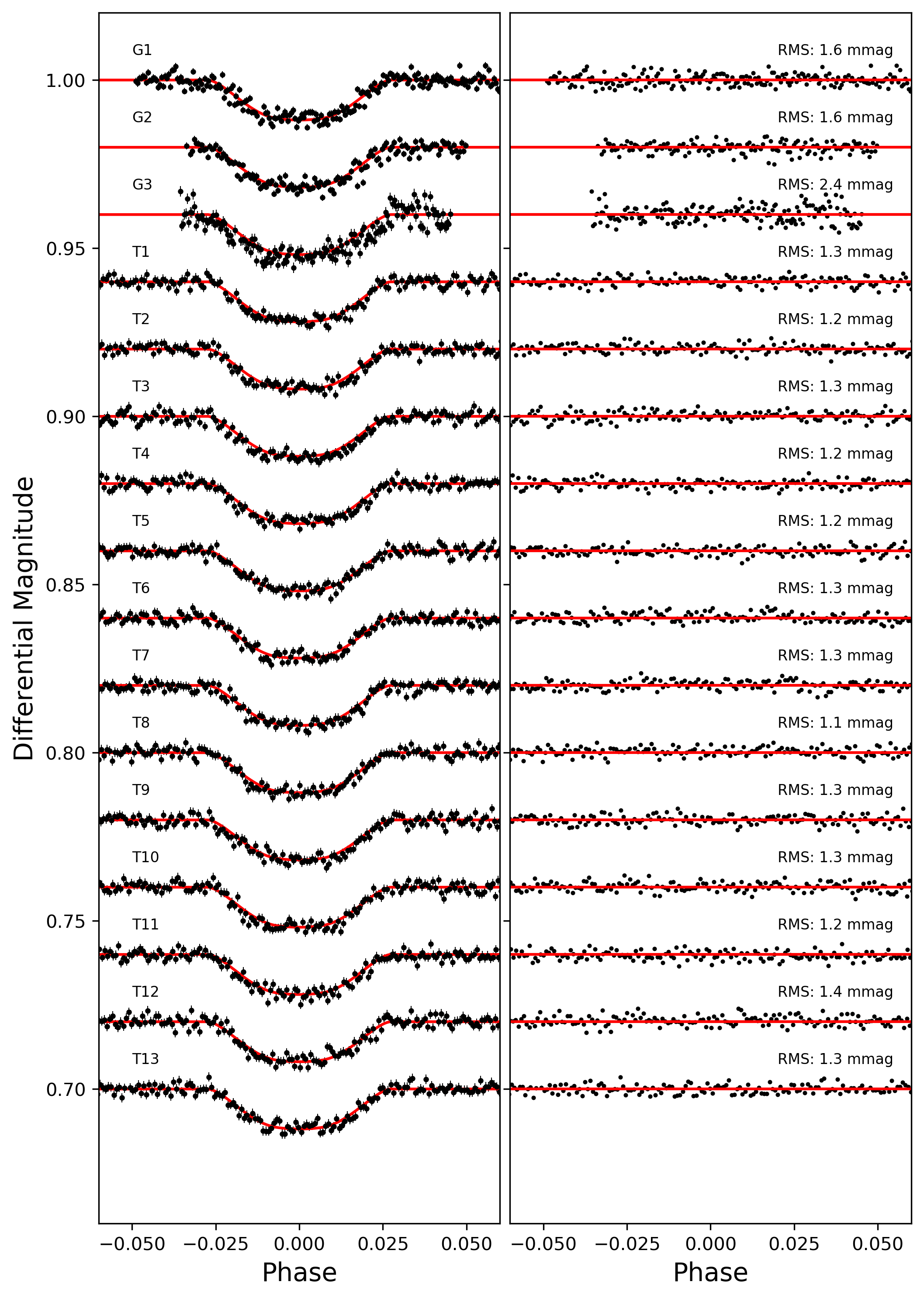}
    \caption{All the observations of WASP-122 b used in this work.}
    \label{WASP-122}
\end{figure}

\begin{figure}
    \centering
    \includegraphics[width = 0.47\textwidth]{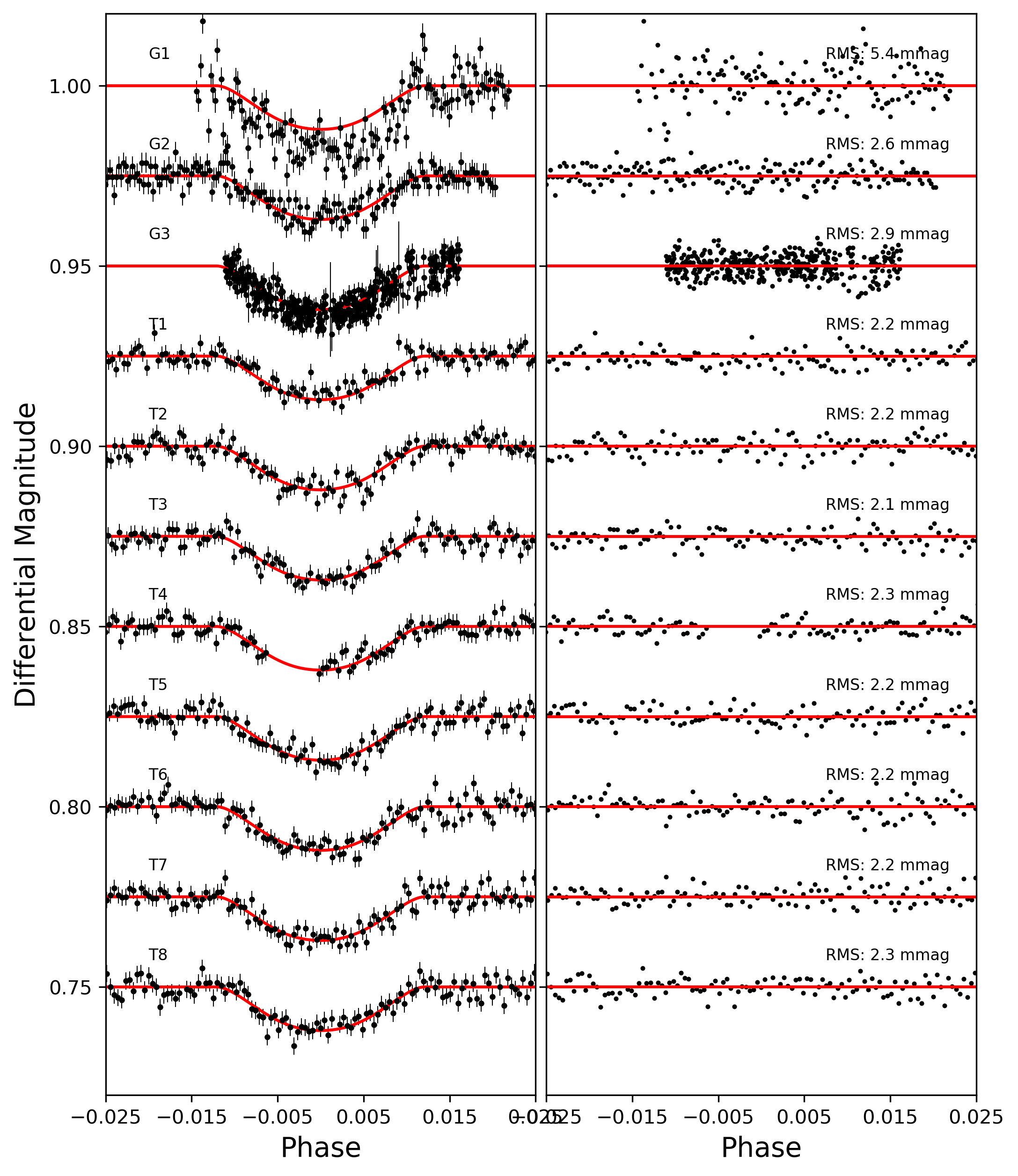}
    \caption{All the observations of WASP-45 b used in this work.}
    \label{WASP-45}
\end{figure}

\begin{figure}
    \centering
    \includegraphics[width = 0.47\textwidth]{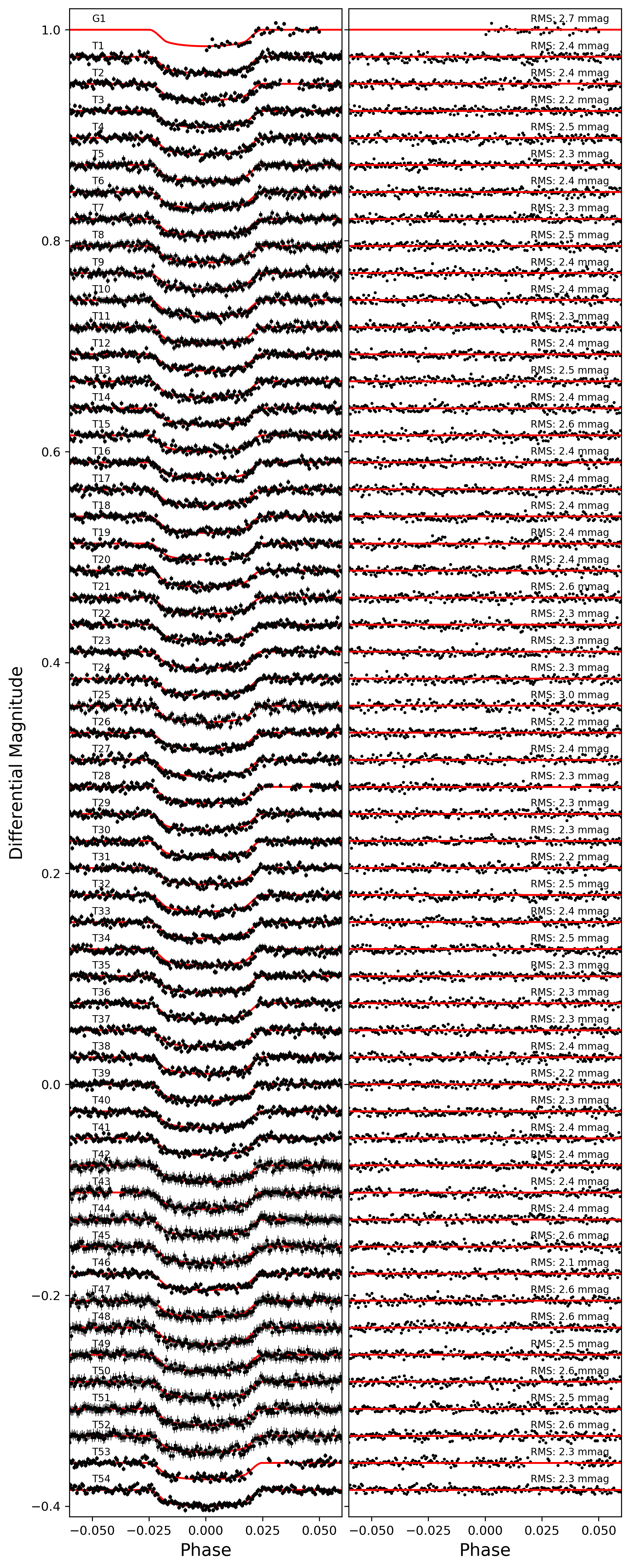}
    \caption{All the observations of WASP-119 b used in this work.}
    \label{WASP_119}
\end{figure}

\begin{figure}
    \centering
    \includegraphics[width = 0.47\textwidth]{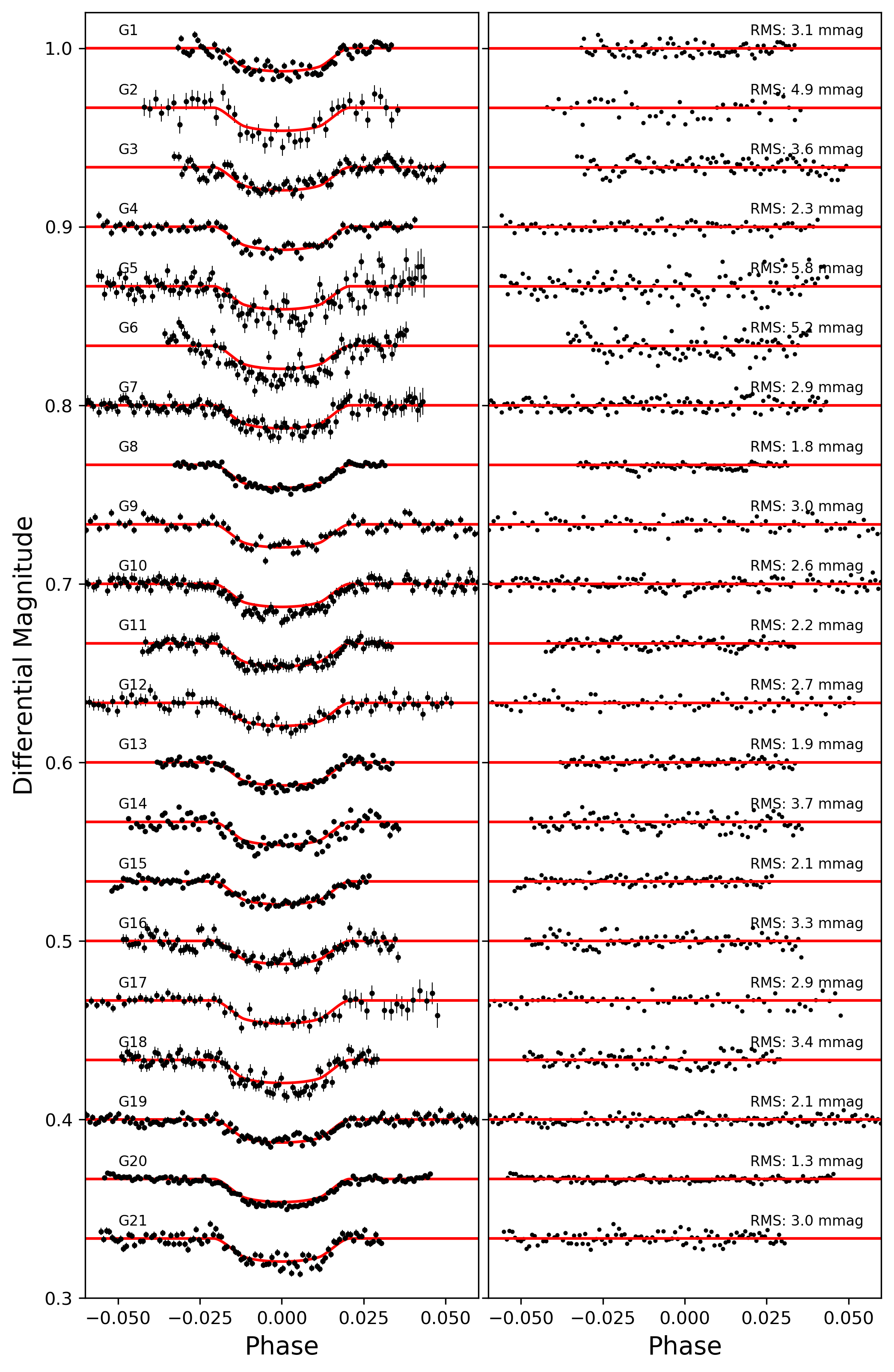}
    \caption{All the observations of KPS-1 b used in this work.}
    \label{KPS-1}
\end{figure}

\begin{figure}
    \centering
    \includegraphics[width = 0.47\textwidth]{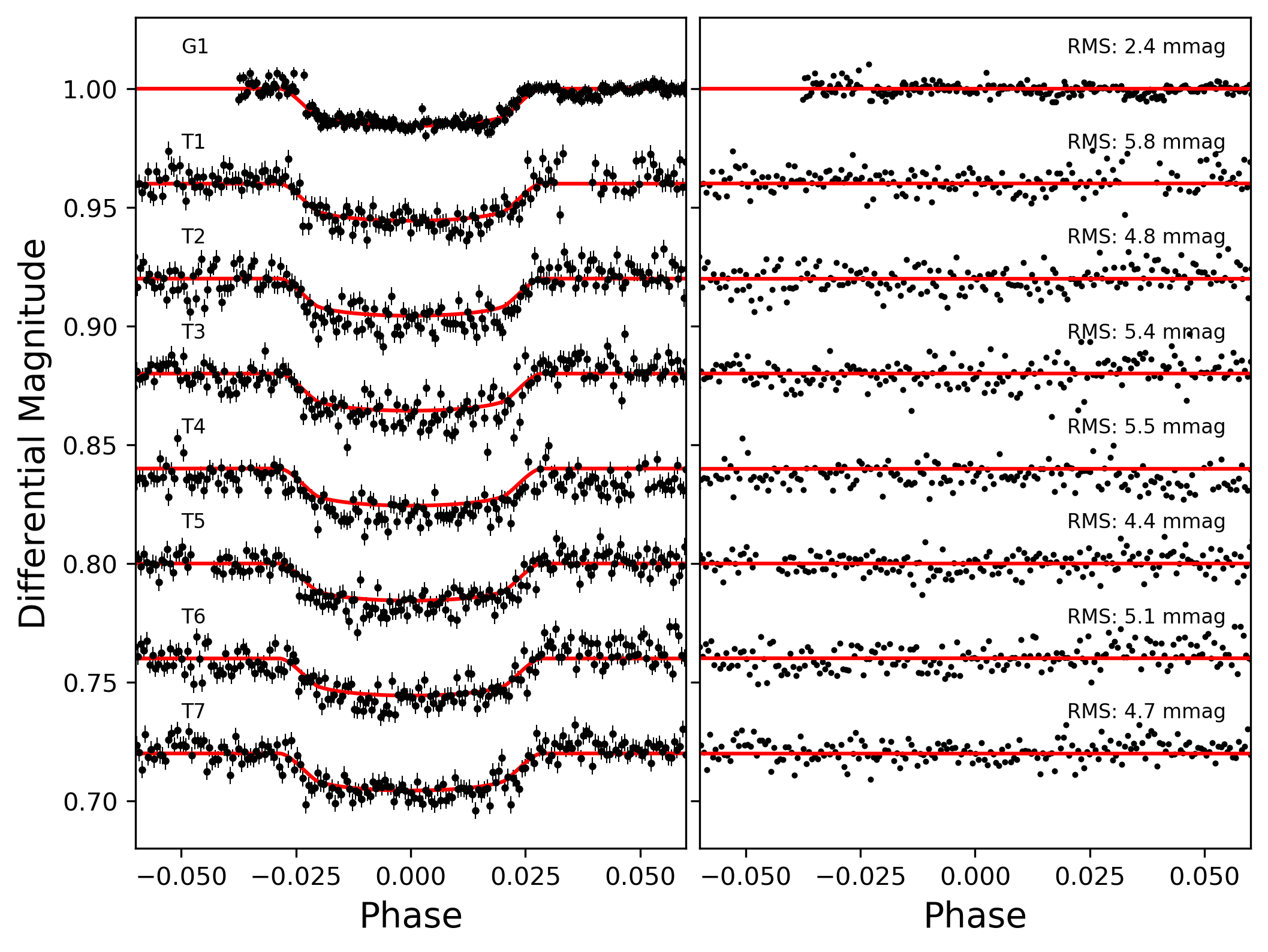}
    \caption{The observation of K2-237 b used in this work.}
    \label{K2-237}
\end{figure}

\begin{figure}
    \centering
    \includegraphics[width = 0.47\textwidth]{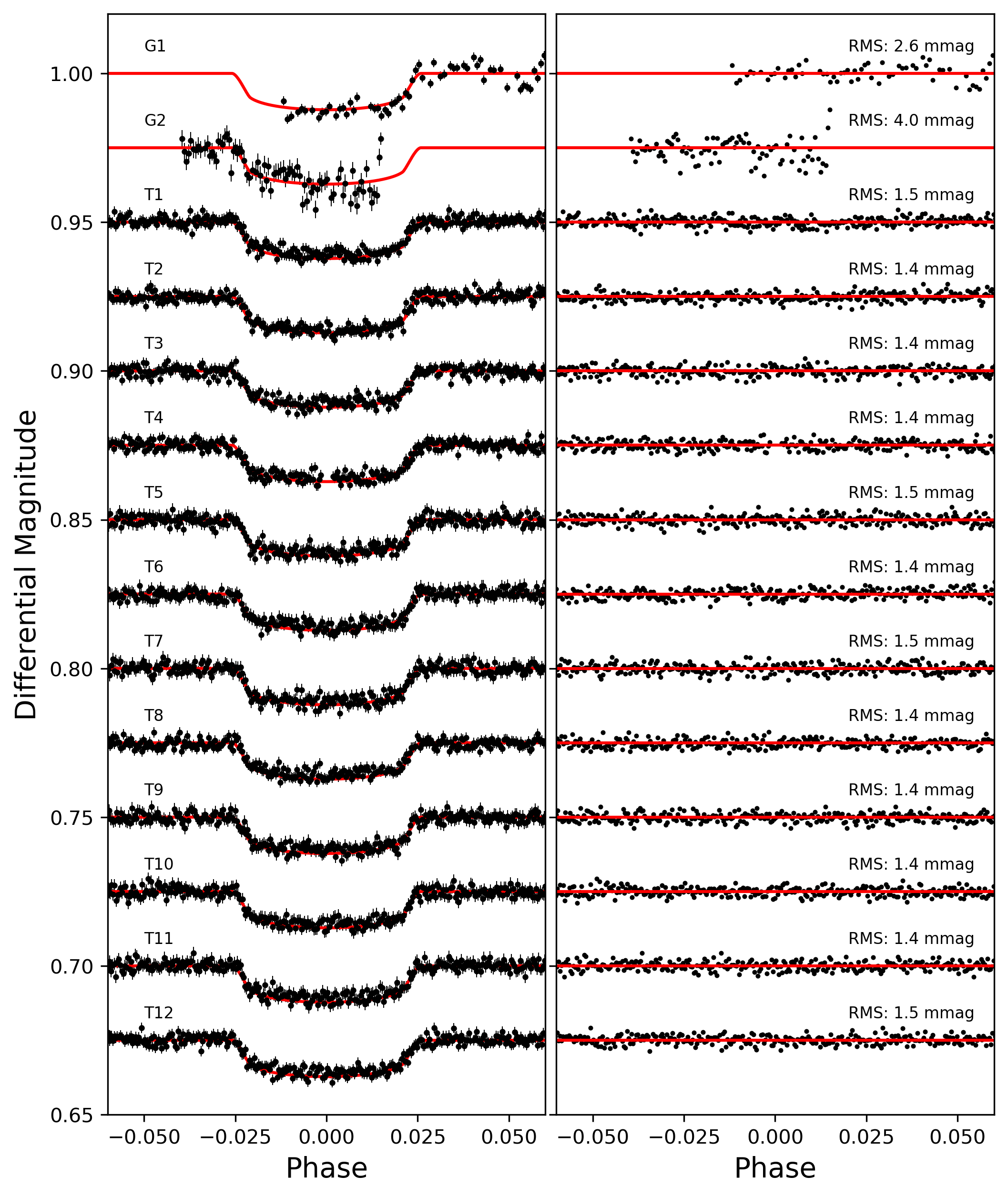}
    \caption{All the observations of KELT-15 b used in this work.}
    \label{KELT-15}
\end{figure}

\begin{figure}
    \centering
    \includegraphics[width = 0.47\textwidth]{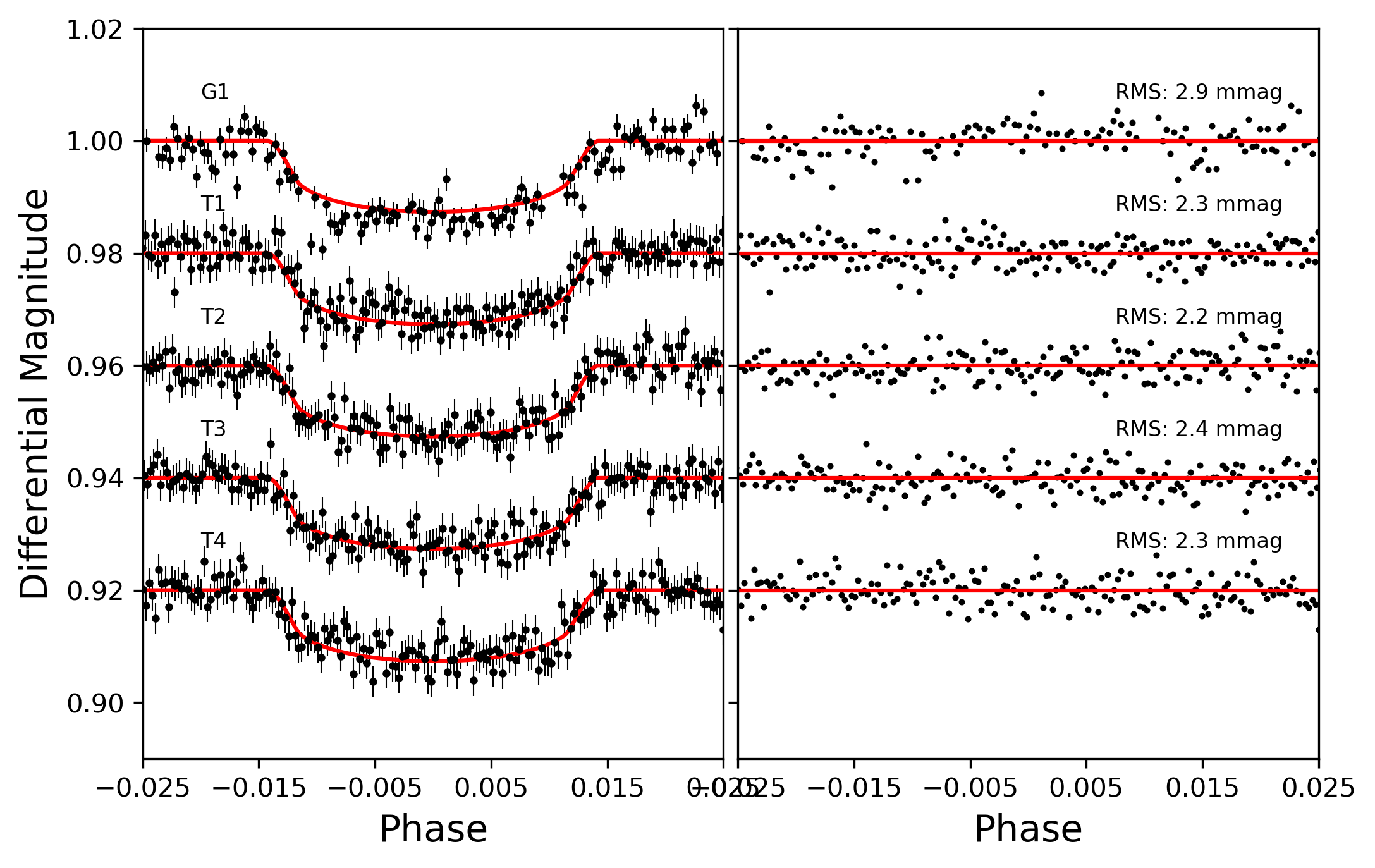}
    \caption{All the observations of WASP-83 b used in this work.}
    \label{WASP_83}
\end{figure}

\begin{figure}
    \centering
    \includegraphics[width = 0.47\textwidth]{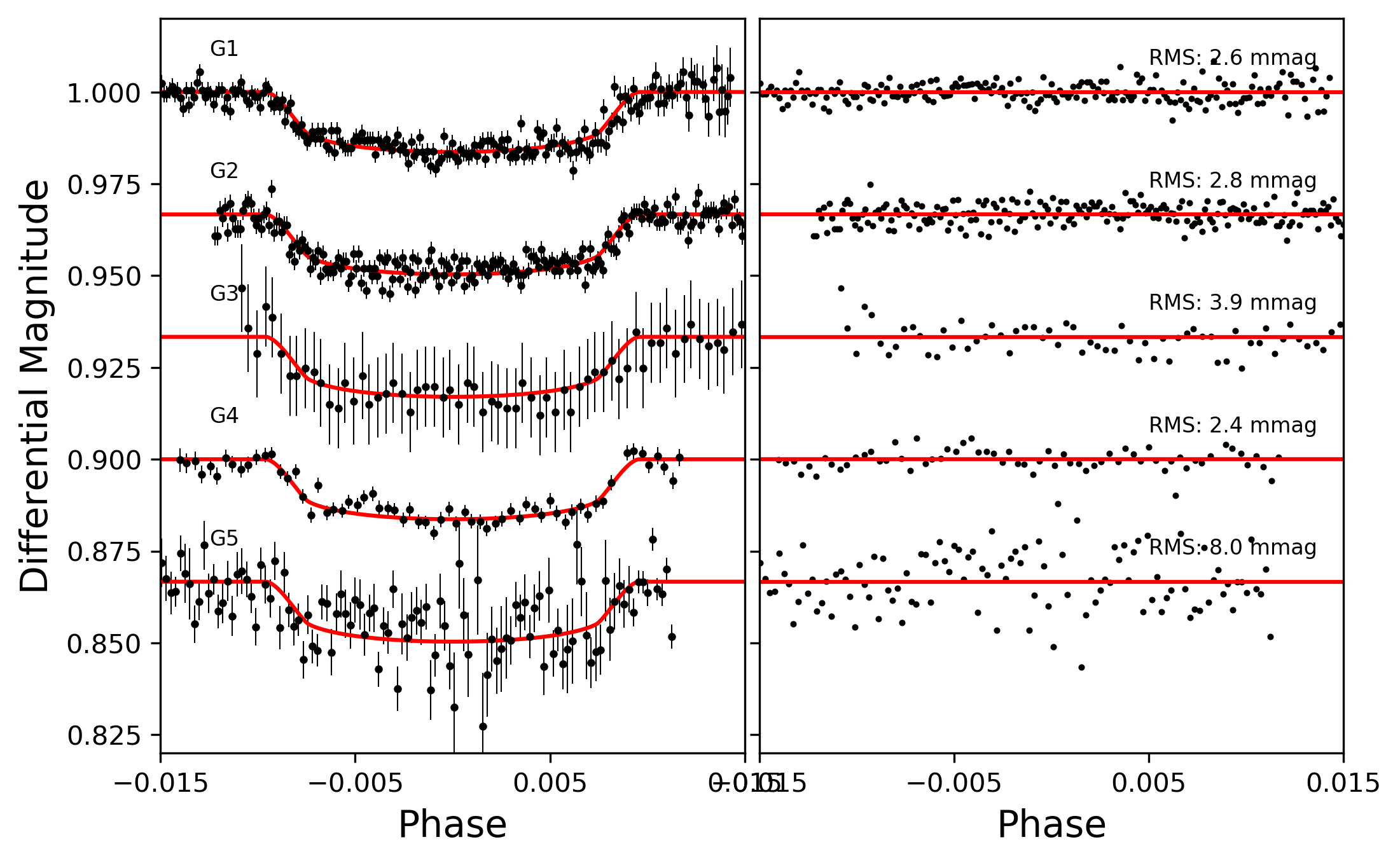}
    \caption{The observation of CoRoT-6 b used in this work.}
    \label{CoRoT6}
\end{figure}

\clearpage

\bibliographystyle{mnras}
\bibliography{main}

\bsp
\label{lastpage}
\end{document}